\documentclass[american, twocolumn]{aastex63}
\usepackage{amsmath, wasysym}
\usepackage{multirow}
\usepackage{scrextend}

\newcommand\plasticc{PLAsTiCC}

\shorttitle{Considerations for optimizing photometric classification of supernovae}
\shortauthors{Alves et al.}

\graphicspath{{./}{figures/}}

\begin{document}

\title{Considerations for optimizing photometric classification of supernovae from the Rubin Observatory}


\correspondingauthor{Catarina S. Alves}
\email{catarina.alves.18@ucl.ac.uk}

\author{Catarina S. Alves}
\affiliation{Department of Physics \& Astronomy, University College London, Gower Street, London WC1E 6BT, UK}

\author{Hiranya V. Peiris}
\affiliation{Department of Physics \& Astronomy, University College London, Gower Street, London WC1E 6BT, UK}
\affiliation{Oskar Klein Centre for Cosmoparticle Physics, Department of Physics, Stockholm University, AlbaNova University Center, Stockholm 10691, Sweden}

\author{Michelle Lochner}
\affiliation{Department of Physics and Astronomy, University of the Western Cape, Bellville, Cape Town, 7535, South Africa}
\affiliation{South African Radio Astronomy Observatory (SARAO), The Park, Park Road, Pinelands, Cape Town 7405, South Africa}
\affiliation{African Institute for Mathematical Sciences, 6 Melrose Road, Muizenberg, 7945, South Africa}

\author{Jason D. McEwen}
\affiliation{Mullard Space Science Laboratory, University College London, Holmbury St Mary, Dorking, Surrey RH5 6NT, UK}

\author{Tarek Allam Jr}
\affiliation{Mullard Space Science Laboratory, University College London, Holmbury St Mary, Dorking, Surrey RH5 6NT, UK}

\author{Rahul Biswas}
\affiliation{Oskar Klein Centre for Cosmoparticle Physics, Department of Physics, Stockholm University, AlbaNova University Center, Stockholm 10691, Sweden}

\author{The LSST Dark Energy Science Collaboration}

\begin{abstract}

The Vera C. Rubin Observatory will increase 
the number of observed supernovae (SNe) by an order of magnitude; 
however, it is impossible to spectroscopically confirm the class for all 
the SNe discovered. Thus, photometric classification is crucial but its 
accuracy depends on the not-yet-finalized observing strategy of Rubin 
Observatory's Legacy Survey of Space and Time (LSST). We quantitatively analyze the impact of the LSST observing strategy on SNe classification using simulated multi-band light curves from the Photometric LSST Astronomical Time-Series Classification Challenge (\plasticc). First, we 
augment the simulated training set to be representative of the 
photometric redshift distribution per supernovae class, the cadence of observations, and the flux uncertainty distribution of the test set. Then we build a classifier using the photometric transient classification 
library \texttt{snmachine}, based on wavelet features obtained from Gaussian process fits, yielding similar performance to the winning 
\plasticc\ entry. We study the classification performance for SNe with different properties within a single simulated observing strategy. We 
find that season length is important, with light curves of $150$ days yielding the highest performance. Cadence also has an important impact on SNe classification; events with median inter-night gap $<3.5$ days yield 
higher classification performance. Interestingly, we find that large gaps ($>10$ days) in light curve observations do not impact performance if sufficient observations are available on either side, due to the 
effectiveness of the Gaussian process interpolation. This analysis is the first exploration of the impact of observing strategy on photometric supernova classification with LSST.

\end{abstract}

\keywords{\href{http://astrothesaurus.org/uat/343}{Cosmology (343)}; \href{http://astrothesaurus.org/uat/1668}{Supernovae (1668)}; \href{http://astrothesaurus.org/uat/1855}{Astronomy software (1855)}; \href{http://astrothesaurus.org/uat/1866}{Open source software (1866)}; \href{http://astrothesaurus.org/uat/1858}{Astronomy data analysis (1858)}; \href{http://astrothesaurus.org/uat/1907}{Classification (1907)}; \href{http://astrothesaurus.org/uat/1954}{Light curve classification~(1954)}}


\section{Introduction} \label{sec:intro}

The upcoming Rubin Observatory Legacy Survey of Space and Time (LSST)
\citep{LSSTScienceBook2009,LSSTObservingStartegy2017,Ivezic2019}
is expected to discover, during its ten-year duration, at least one order of magnitude more supernovae (SNe) than the current
available SNe samples \citep{Guillochon2017}. Traditionally, SNe that are used in astrophysical and cosmological studies need to be spectroscopically classified \citep[e.g. ][]{Riess1998,Astier2006,Kessler2009}. However, this will be impossible for most events detected by LSST due to the limited spectroscopic resources; thus, LSST will rely on photometric classification, using the events that will be spectroscopically classified as its training set.

Previous efforts to understand the strengths and limitations of photometric
classification algorithms resulted in the Supernova Photometric Classification Challenge
 \citep[\texttt{SNPhotCC};][]{Kessler2010spcc} in preparation for the Dark Energy Survey
\citep[DES;][]{DES2005}. Recently, the Photometric LSST Astronomical
Time-Series Classification Challenge\footnote{\url{https://www.kaggle.com/c/PLAsTiCC-2018/}}
\citep[PLAsTiCC;][]{allam2018photometric,Kessler2019} was launched in preparation
for LSST, which will reach fainter magnitudes and have a $\sim4$
times larger survey area compared to DES. The classifiers applied to the datasets from these challenges employed parametric
fits, template fits, and machine learning models such as neural networks,
boosted decision trees, support vector machine, and gradient boosting
\citep[e.g.][]{Kessler2010results,Lochner2016,Charnock2017,Pasquet2019,Muthukrishna2019,Villar2020}. 

To obtain accurate classification, the training set must be representative of the test set (e.g. \citealt{Lochner2016}). However, photometric
classifiers are typically trained with non-representative spectroscopically-confirmed
events that are biased towards lower redshifts. Thus, recent work has
focused on overcoming the lack of representativeness \citep{Muthukrishna2019, Pasquet2019, Revsbech2017, Boone2019, Carrick2021}. Photometric classification performance also depends on the survey observing strategy; however, this dependence has not yet been explored.

The LSST observing strategy encompasses diverse considerations such as season length, survey footprint, single visit exposure time, inter-night gaps, and cadence of repeat visits in different passbands. The observing strategy is currently being optimized \citep{LSSTObservingStartegy2017,ivezic2018call,DESCObservingStrategy2018,Gonzalez2018,Laine2018,Jones2020}, a challenging task since the survey has diverse goals \citep{LSSTScienceBook2009,Ivezic2019}.
Recently, the Rubin Observatory LSST Dark Energy Science Collaboration
(DESC) Observing Strategy Working Group investigated the impact of observing 
strategy on cosmology and made recommendations for its optimization
\citep{DESCObservingStrategy2018ddf,DESCObservingStrategy2018,osmetrics}.
In particular, SNe cosmology requires a high and regular cadence
with long season lengths (how long a field is observable in a year). 

In this work, we upgrade the photometric transient classification
library \texttt{snmachine}\footnote{\url{https://github.com/LSSTDESC/snmachine}\label{fn:snmachine}}
\citep{Lochner2016} for use with LSST data and build a classifier based on wavelet features obtained from Gaussian process (GP) fits. We also include the host-galaxy
photometric redshifts and their uncertainties as features.
We make several other improvements to deal with the greater realism of the \plasticc\ data, including training set augmentation. Using this improved classifier we study the performance of photometric SNe classification for subsets of light curves with different cadence properties, using the single observing strategy simulated for the \plasticc\ challenge. We note that this approach is different from studying the classification performance for different observing strategies with fixed total exposure time, where a reduced season length could be compensated by a higher cadence.

In Sections \ref{sec:data} and \ref{sec:methods} we summarize the
\plasticc\ dataset and describe the classification pipeline, respectively.
Section \ref{subsec:Augmentation} focuses on the augmentation methodology. Our results and their implications for observing strategy are described in Section \ref{sec:results}. We conclude in Section \ref{sec:conclusion}.


\section{\plasticc\ Dataset} \label{sec:data}

The \plasticc\ \citep{allam2018photometric,PLASTICCTeam2019} dataset
consists of simulations of $18$ different classes of transients and
variable stars. It contains three-year-long light curves of $3.5$
millions events observed in the LSST $ugrizy$ passbands, as well as their host-galaxy photometric redshifts and uncertainties. Although
the simulations included realistic observing conditions, the observing strategy used\footnote{Simulation \texttt{minion\_1016}: \url{https://docushare.lsst.org/docushare/dsweb/View/Collection-4604}}
is now outdated \citep{Jones2020}. \plasticc\ mimicked future LSST
observations in two survey modes: the Wide-Fast-Deep (WFD) survey,
which covers almost half the sky and was used for $99\%$ of the events,
and the Deep-Drilling-Fields (DDF) survey, small patches of the sky
with more frequent and deeper observations that have smaller flux
uncertainties. 

The simulations were divided into a non-representative spectroscopically-confirmed
training set biased towards brighter events, and which is $0.2\%$ of the size of the test set. The training set was 
much smaller, to mimic the data that will be available at the start of 
LSST science operations from current and near-term spectroscopic surveys.
In particular, the training set was loosely modeled on the magnitude-limited 4-metre Multi-Object Spectroscopic Telescope Time Domain Extragalactic Survey \citep{Swann2019}, resulting in a sample with a mean redshift $\sim0.3$.
The unblinded dataset is available
in \citet{PLASTICCTeam2019}, the model libraries are presented in \citet{PLAsTiCCModelers2019}, and more details about  the models and
simulations, including the description of the training set, host-galaxy
photometric redshifts and their uncertainties, are given  in 
\citet{Kessler2019}. In this work we provide observing strategy recommendations to improve photometric classification of SNe in particular, so we restrict ourselves to the \mbox{\plasticc\ }classes SN Ia, SN Ibc, and SN II; Table \ref{tab:Summary-objs-used} shows a breakdown of the numbers of SNe in each class.


\section{Classification pipeline} \label{sec:methods}

In this section we describe how we upgraded the photometric classification pipeline \texttt{snmachine} for use with \plasticc\ data. Augmentation was a crucial step in this process, and it is discussed in greater detail in Section~\ref{subsec:Augmentation}.

\subsection{Light Curve Preprocessing \label{subsec:Preprocess}}

\begin{figure}
\begin{centering}
\includegraphics[width=0.4\textwidth]{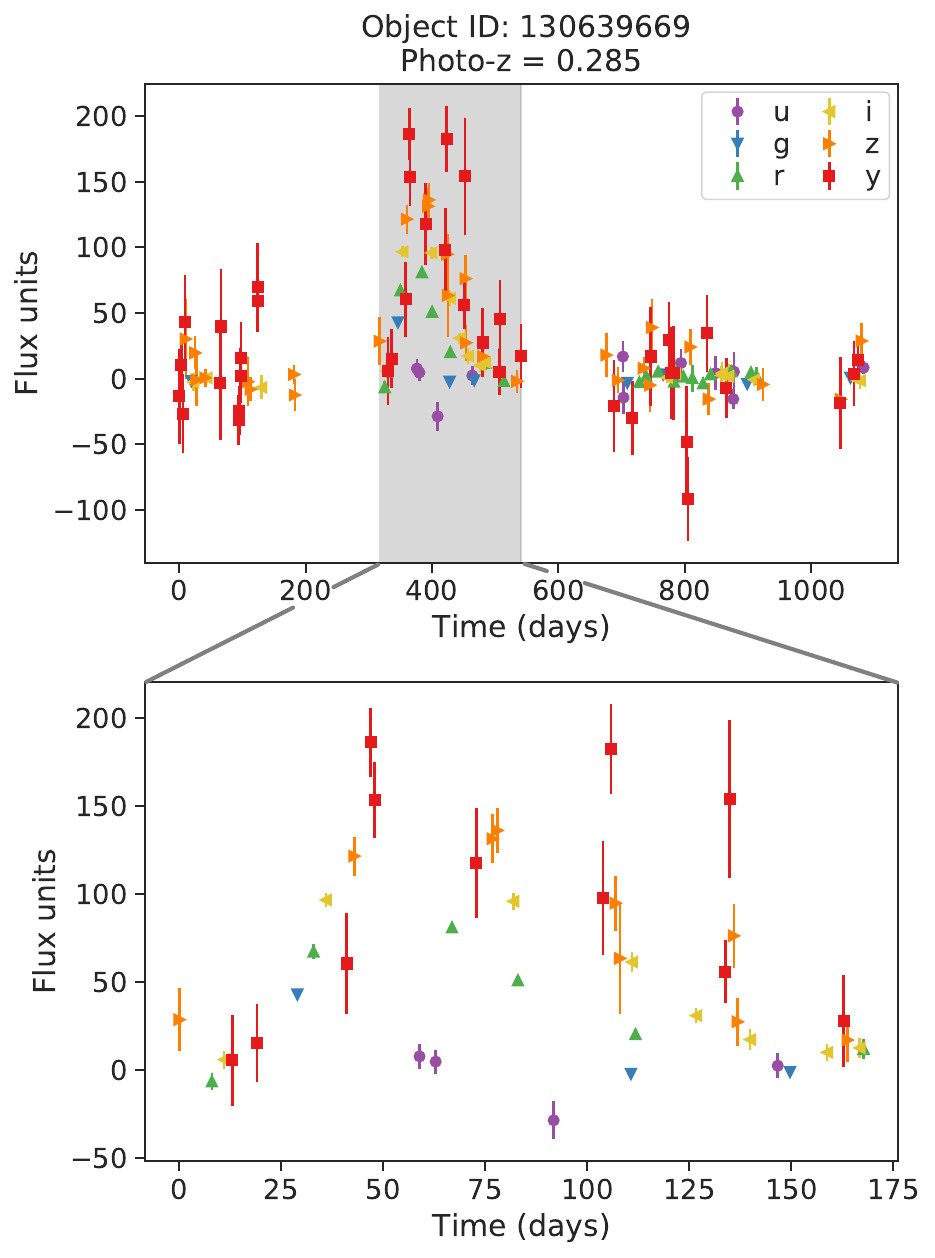}
\par\end{centering}
\caption{Example of a simulated LSST type Ibc supernova light curve from \plasticc, showing how we preprocess light curves to remove season gaps. The original
and processed light curves are shown in the top and bottom panels,
respectively. The processed light curve corresponds to the shaded
region on the original light curve with the first observation translated
to time zero. The observations in different passbands are shown in
different colors. \label{fig:gapless50 before after}}
\end{figure}

PLAsTiCC light curves have long gaps ($>50$ days) in the observations
because any given sky location is not visible from the Vera C. Rubin
Observatory site for several months of the year.
Additionally, the SNe are only detected for a few months so including
the entire three-year-long light curve provides irrelevant information
to the classifier, which in turn degrades its performance. In order
to isolate the observing season that contains the SNe, we selected the season which contains the observations flagged as detected, and which has no inter-night gaps larger than $50$ days. To introduce uniformity in the
dataset, we translated the resulting light curves so their first 
observation is at time zero. However, this results in light curves that peak at different times, so we explored additionally shifting all training set 
light curves randomly in time to capture a larger variability of 
peak times. We found that augmenting with this random shift led to a less 
representative training set, and thus to a worse classification performance. Therefore, in this work, we simply aligned the first observation of the 
training events at time zero, such as we did for the test set. Figure \ref{fig:gapless50 before after} shows
 an example of light curve preprocessing.

\begin{table}
\caption{Breakdown of the number of SNe per class used in this work (see simulation details in \citet{Kessler2019}). For each class, the number of events in the training and test set is shown.
\label{tab:Summary-objs-used} }

\begin{centering}
\smallskip{}

\begin{tabular}{ccc}
\hline
SN class & $N_{\mathrm{training}}$ (\%) & $N_{\mathrm{test}}$ (\%)\tabularnewline
\hline 
SN Ia & $2\,313$ ($58\%$) & $1\,659\,831$ ($59\%$)\tabularnewline
SN Ibc & $484$ ($12\%$) & $175\,094$ ($6\%$)\tabularnewline
SN II & $1\,193$ ($30\%$) & $1\,000\,150$ ($35\%$)\tabularnewline
\hline 
Total & $3\,990$ ($100\%$) & $2\,835\,075$ ($100\%$)\tabularnewline
\hline
\end{tabular}
\par\end{centering}
\end{table}

\subsection{Gaussian Process Modeling of Light Curves\label{subsec:Model-light-curves}}

\begin{figure*}
\begin{centering}
\includegraphics[width=0.4\textwidth]{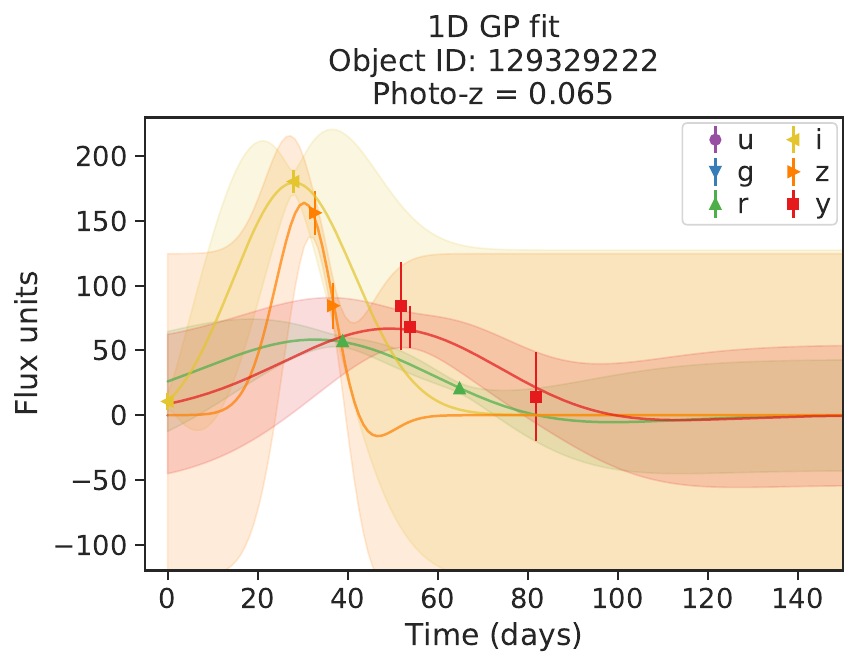}\includegraphics[width=0.4\textwidth]{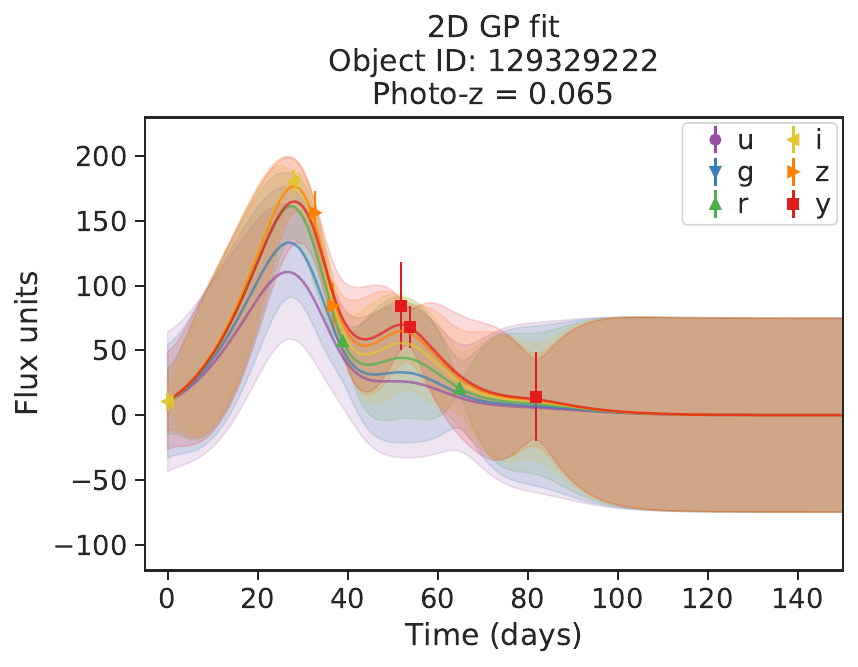}
\par\end{centering}
\caption{SN Ibc light curve, where the points show the observations, along with their errorbars, and
the lines and the shaded regions show the mean and standard
deviation of the GP fit, respectively. The left panel shows the one-dimensional GP fit to each
available passband and the right panel shows the two-dimensional GP
fit to all the passbands (shown in different colors).
The two-dimensional GP infers the light curve in passbands where there
are no (or few) observations, unlike the one-dimensional GP.
\label{fig:example 2d GP}}
\end{figure*}

We modeled each light curve with a GP
regression \citep[e.g.][]{MacKay2003,Rasmussen2005}, following previous works that successfully used GP-modeled light
curves in their classification pipelines \citep[e.g. ][]{Lochner2016,Revsbech2017}.
Unlike the previous examples that fitted separate GPs to each passband,
\citet{Boone2019} fitted GPs both in time and wavelength, thus allowing
the GPs to incorporate cross-band information. Figure \ref{fig:example 2d GP}
shows that such a two-dimensional GP fit infers the SNe light curve
even in passbands where there are none or only a few observations,
in contrast to the one-dimensional GP fit. Thus, we used two-dimensional GPs to fit light curves both in time
and wavelength.

We chose a null mean function for the GP, modeling the events as perturbations
to a flat background. Following \citet{Boone2019}, we used the once-differentiable
Matérn 3/2 kernel for the GP covariance, which is appropriate for
modeling explosive transients with sudden changes in their flux. The
time dimension length-scale and amplitude were optimized per event, using maximum likelihood estimation. We fixed the length-scale of the wavelength dimension to $6000\,\mathring{\mathrm{A}}$ as in \citet{Boone2019}, since they found that this value produces reasonable models for all classes in \plasticc. The GPs were implemented with the package \texttt{George}\footnote{\url{george.readthedocs.io/}}
\citep{george2014}. 

\subsection{Feature Extraction\label{subsec:Extract-wavelet-features}}

In this work we followed the wavelet decomposition
approach of \citet{Lochner2016} to extract features. Since this is a model-independent approach to feature extraction, it does not assume
any physical knowledge about the observed phenomena; hence it is applicable
to any time-series data. Moreover, recent results showed wavelet decomposition
was successful for general transient classification \citep{Varughese2015,Lochner2016,Sooknunan2021,Narayan2018}. This model-independent approach had not been used previously by the winning \plasticc\ entries.

Following \citet{Lochner2016}, we used a Stationary Wavelet Transform
and the \texttt{symlet} family of wavelets; the wavelet decomposition was implemented with the package \texttt{PyWavelets} \citep{Lee2019}. To obtain the wavelet decomposition, we first used the GPs to interpolate all light curves onto the same time grid of $277$ days (maximum light
curve length of the events); we chose approximately one grid point per day and used a two-level wavelet decomposition, following \citet{Lochner2016}.
These choices resulted in $6624$ (highly redundant) wavelet 
coefficients per event.
While it is common to combine GP fits and wavelet analysis \citep[e.g.][and references therein]{Chen2013,Istas1992,Pope2019}, we note that our method of modeling the sparse light curves with GP fits and then using wavelet decomposition to obtain classification features is unusual. This approach was briefly mentioned in \citet{Varughese2015}, and firstly implemented in \citet{Lochner2016}.

Following \citet{Lochner2016}, we reduced the dimensionality
of this wavelet space using Principal Component Analysis (PCA) \citep{Pearson1901,Hotelling1933}
on the wavelet coefficients of the augmented training set. After comparing the classifier performance on a validation set (we set aside $5\%$ of the test set for validation) with different numbers of PCA components,
we found that $20-50$ components were the best to distinguish different
types of SNe (their log-loss differs by around $2\%$); we chose $40$ components ($99.995\%$ of the total variance)
due to its slightly better performance. 

Finally, we also include the photometric redshift and its uncertainty as classification features. Unlike our previous results on the SNPhotCC challenge \citep{Lochner2016} we find that these features are crucial for solving the more realistic classification challenge presented by the \plasticc\ data. This is also confirmed by other \plasticc\ analyses \citep{Boone2019, ResultsPlasticc2020}. 

\subsection{Classification\label{subsec:Train-GBDT-classifier}}

We augmented the training set as described in Section~\ref{subsec:Augmentation}, prior to training a classifier. We used the Gradient Boosting Model implementation of the package \texttt{LightGBM}\footnote{\url{lightgbm.readthedocs.io}} \citep{ke2017lightgbm}, in particular
the Gradient Boosting Decision Tree (GBDT) \citep{Friedman2001}. These are ensemble classifiers that produce predictions using ensembles of decision trees. The boosting improves the ensemble prediction by sequentially adding new decision trees that prioritize difficult-to-classify events. Boosted decision trees are commonly used in machine learning pipelines, including most of the top solutions to PLAsTiCC challenge \citep{ResultsPlasticc2020}, due to their robust predictions, capacity for handling missing data, and flexibility \citep{Friedman2001, ke2017lightgbm}.

We optimized the GBDT hyperparameters (parameters of the model that must be set before the learning process starts) by maximizing the performance
of a $5$-fold cross-validated grid-search on the augmented training
set. First, each hyperparameter was optimized individually using a one-dimensional grid, keeping the other hyperparameters at default values. Then, we constructed a six-dimensional grid with three possible values for each hyperparameter informed by the earlier one-dimensional optimization. Finally we optimized this six-dimensional grid through a standard grid search. The resulting hyperparameter
values are shown in Table \ref{tab:Hyper-parameters}. Since training and testing on the same events leads to overfitting, we placed in the same cross-validation fold all synthetic events that were derived from the same original event. While alternative hyperparameter optimization techniques can be considered
\citep[e.g. Bayesian optimization;][]{Mockus1978,snoek2012}, a simple
grid search strategy as described above proved to be effective.

\begin{table}
\caption{Optimized hyperparameter values used for the \texttt{LightGBM} model.
A description of the hyperparameters is given in the \protect\href{https://lightgbm.readthedocs.io/en/latest/pythonapi/lightgbm.LGBMClassifier.html\#lightgbm.LGBMClassifier}{library documentation}.\label{tab:Hyper-parameters}}

\centering{}
\begin{tabular}{ccc}
\hline
Hyperparameter   & WFD setting & DDF setting\tabularnewline
\hline 
\texttt{boosting\_type} & \texttt{gbdt} & \texttt{gbdt}\tabularnewline
\texttt{learning\_rate} & $0.24$ & $0.24$\tabularnewline
\texttt{max\_depth} & $16$ & $19$\tabularnewline
\texttt{min\_child\_samples} & $25$ & $70$\tabularnewline
\texttt{min\_split\_gain} & $0.3$ & $0.3$\tabularnewline
\texttt{n\_estimators} & $115$ & $45$\tabularnewline
\texttt{num\_leaves} & $50$ & $50$\tabularnewline
\hline
\end{tabular}
\end{table}

\subsubsection{Performance Evaluation\label{subsec:Performance-metric}}

In order to evaluate the classification performance, we used the PLAsTiCC weighted log-loss metric \citep{allam2018photometric,malz2018metric} given by
\begin{equation}
\text{Log-loss}=-\left(\dfrac{\sum_{i=1}^{M}w_{i}\cdot\sum_{j=1}^{N_{i}}\frac{y_{ij}^{*}}{N_{i}}\cdot\ln p_{ij}}{\sum_{i=1}^{M}w_{i}}\right)\,,\label{eq:logloss plasticc}
\end{equation}
 where $M$ is the total number of classes, $N_{i}$ is the number
of events in class $i$, $y_{ij}^{*}$ is $1$ if observation $j$
belongs to type $i$ and $0$ otherwise, $p_{ij}$ is the predicted
probability that event $j$ belongs to class $i$ and $w_{i}$ is
the weight of the class $i$. The weights can be changed to give different
importances to different classes; however, following the PLAsTiCC challenge, we gave the same weight to every SNe class. 

We used confusion matrices to visualize the mislabeled classes; Table
\ref{tab:Confusion-matrix-scheme} shows the confusion matrix for
a binary classification. For ease of comparison, we normalized the
confusion matrices by dividing each entry by the true number of each
SNe class; hence the identity matrix represents a perfect classification.

\begin{table}
\caption{Confusion matrix for binary classification.\label{tab:Confusion-matrix-scheme}}

\centering{}%
\begin{tabular}{cc|cc}
\hline
 &  & \multicolumn{2}{c}{True class}\tabularnewline
 &  & Positive (P) & Negative (N)\tabularnewline
\hline 
Predicted  & P & True positive ($\mathrm{TP}$) & False positive ($\mathrm{FP}$)\tabularnewline
class & N & False negative ($\mathrm{FN}$) & True negative ($\mathrm{TN}$)\tabularnewline
\hline
\end{tabular}
\end{table}

For a single SNe class, it is also common to use the recall (also
called completeness/sensitivity) to measure the fraction of correctly-classified SNe, and the precision to measure the fraction of SNe assigned
to the considered class that are indeed from that class. These are defined as 
\begin{equation}
\mathrm{recall}=\dfrac{\mathrm{TP}}{\mathrm{TP}+\mathrm{FN}}
\end{equation}
 and 
\begin{equation}
\mathrm{precision}=\dfrac{\mathrm{TP}}{\mathrm{TP}+\mathrm{FP}}\,.
\end{equation}

\begin{figure*}
\begin{centering}
Redshift distribution per class
\par\end{centering}
\begin{centering}
\includegraphics[width=0.3\textwidth]{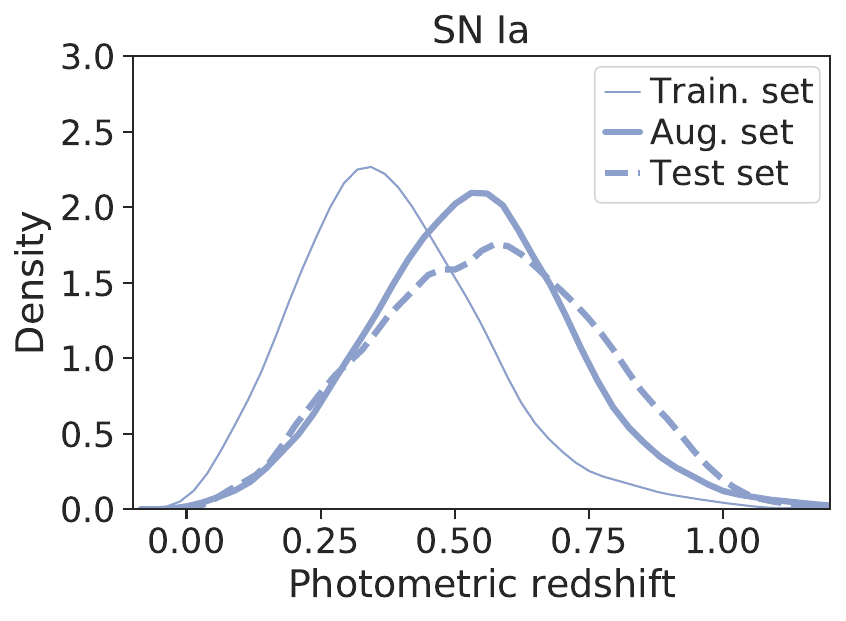}\includegraphics[width=0.3\textwidth]{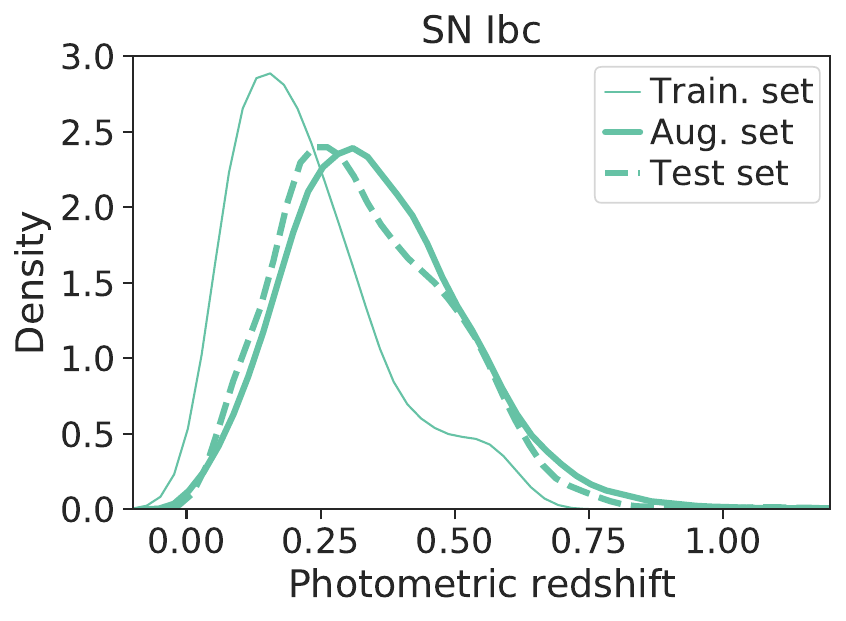}\includegraphics[width=0.3\textwidth]{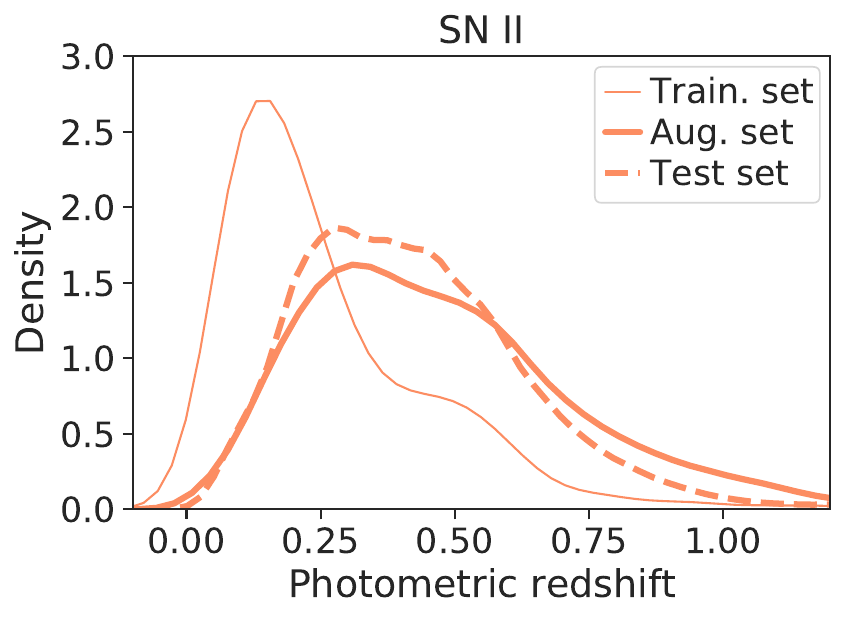}
\par\end{centering}
\caption{Host galaxy photometric redshift distribution per supernova class,
where SN Ia, SN Ibc, SN II are shown, respectively, on the left, middle
and right panels. The distribution of the training, augmented training
and test sets are shown as fine solid, bold solid and dashed lines,
respectively. Although the training set distribution is not representative
of the test set, the augmented training set (Section \ref{subsec:Augmentation}) is close to the desired test distribution.
\label{fig:z class distr}}
\end{figure*}

\begin{figure*}
\begin{centering}
\includegraphics[width=0.4\textwidth]{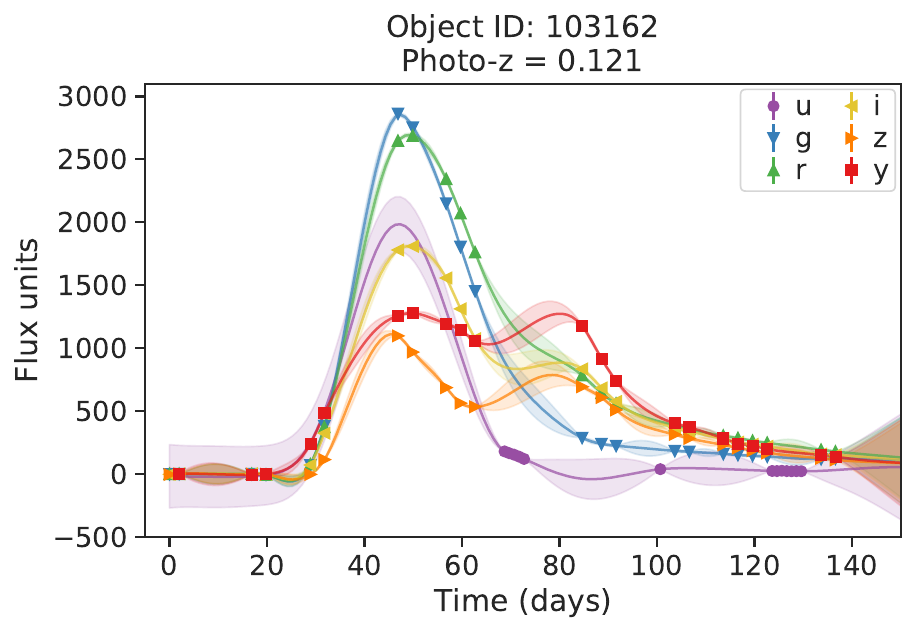}\includegraphics[width=0.4\textwidth]{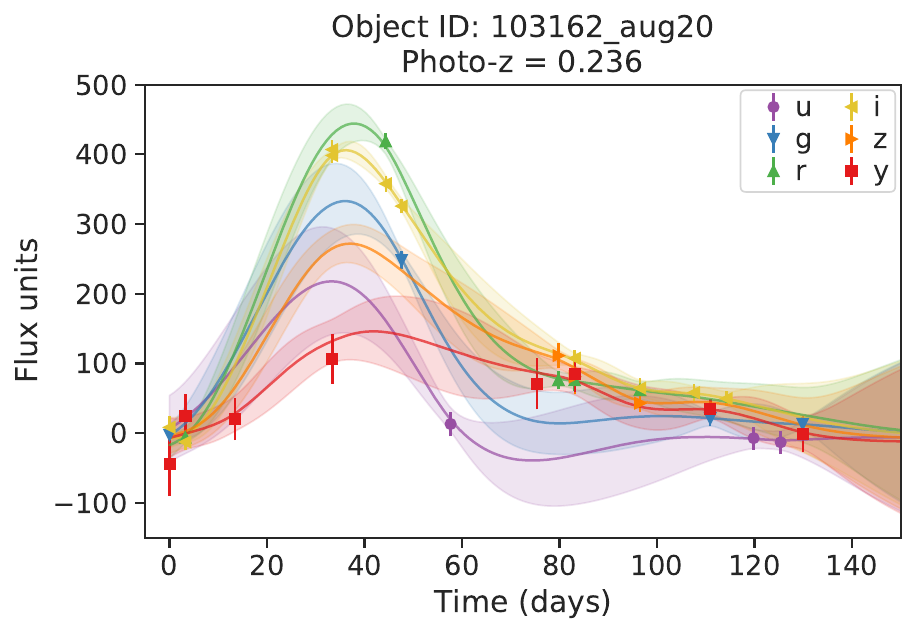}
\par\end{centering}
\caption{The left panel shows a SN Ibc light curve where the points show the observations, along with their errorbars, and the lines and the shaded regions show the mean and standard deviation of the GP fit, respectively. The right panel shows a
synthetic event at a different redshift generated from the original
event using the procedure described in Section \ref{subsec:Augmentation}. Note that, in this example, the original event was simulated in the higher-cadence DDF survey and we generated the synthetic event in the WFD survey.\label{fig:example aug}}
\end{figure*}


\section{Augmentation \label{subsec:Augmentation}}

As previously outlined, the PLAsTiCC training set is non-representative of the test set in redshift (see Figure~\ref{fig:z class distr}) and also imbalanced: the most common SNe class has $\sim4.8$
times more events than the least common. However, to obtain accurate
classification the training set must be representative \citep{Lochner2016}
and balanced (as later discussed in Section \ref{subsec:Number-and-Distribution}).

Recent augmentation approaches rely on generating synthetic light curves
from the GPs fitted to training set events \citep{Revsbech2017,Boone2019}.
In particular, \citet{Boone2019} simulated new sets of observations
for each object such that they match the cadence, depth and uncertainty
of observations of the test set, which ensured the representativity
of these properties irrespective of the quality of the original
event. The augmented observations were drawn from the mean prediction
of the GP, and blocks of observations were dropped to simulate season
boundaries. Additionally, \citet{Boone2019} introduced redshift augmentation, where the observations of a new synthetic event are simulated at a different redshift from the original. 

We adapted the approach used in \citet{Boone2019} for our training set augmentation. Figure \ref{fig:example aug} shows a synthetic light curve generated using our augmentation procedure, which can be summarized as follows:
\begin{enumerate}
\item Choose the number of synthetic events to create (Section \ref{subsec:Number-and-Distribution}).
\item Model the original light curve with a two-dimensional GP fit in time
and wavelength (as described in Section \ref{subsec:Model-light-curves}).
\item Choose a redshift for the synthetic event (Section \ref{subsec:redshift_augmentation}).
\item Create synthetic observations at the new redshift, making use of the GP fit to the original event (Section \ref{subsec:generating_synthetic}).
\item Generate a photometric redshift (Section \ref{subsec:photometric_redshift}). 
\end{enumerate}

The WFD and DDF surveys have very different characteristics and enable qualitatively different science goals. Hence we found it is important to use customized augmentation for the two survey-modes, in contrast to the approach of the winning \plasticc\ entries. Since the DDF survey has a different redshift distribution, higher cadence, and higher signal-to-noise ratio than the WFD survey, we must use a different augmentation and, consequently, a different classifier. 

We now describe the augmentation procedure in detail. The reader should keep in mind, where relevant, that the augmentation procedure was customized for the two survey modes as necessary.


\subsection{Number and Class Balance of Synthetic Events \label{subsec:Number-and-Distribution}}

As we wish to optimize classification performance for all SNe classes, we generated an augmented training set with the same number of events per class (i.e., a balanced training set). We also investigated an augmentation of the training set to resemble the class proportions of the test set ($\sim59\%$ SN Ia, $\sim6\%$ SN Ibc, $\sim35\%$ SN II). However this gave worse performance, biasing the predictions toward the most common class.

We determined that the performance of the classifier stabilized when the size of the WFD augmented training set was around $4\times10^{4}$.
In this final configuration, each training set SN was augmented up to $140$ times. The DDF augmented training set stabilized around $8\times10^{3}$, and each DDF training set SN was augmented up to $70$ times.

\subsection{Redshift Augmentation}
\label{subsec:redshift_augmentation}

As previously outlined, redshift augmentation was found to be critical for the \plasticc\ dataset \citep{Boone2019}. Figure
\ref{fig:z class distr} shows the bias of the training set towards low-redshift events in comparison to the test set. We augmented each training set event of the WFD survey between
\begin{equation}
\begin{aligned}z_{\mathrm{min}} & \approx\max\left\{ 0,0.90\,z_{\mathrm{ori}}-0.10\right\} \,\text{and} \\
z_{\mathrm{max}} & \approx1.43\,z_{\mathrm{ori}}+0.43 \, ,
\end{aligned}
\,\label{eq:redshift range z-max-scale mini}
\end{equation}
where $z_{\mathrm{ori}}$ is the spectroscopic redshift of the original event. For the augmentation, we used a target distribution that is \emph{class-agnostic}. 
First, we drew an auxiliary value $z^{*}$ from a log-triangular distribution with minimum value and mode $\log(z_\mathrm{min})$, and maximum value $\log(z_\mathrm{max})$. Then, we calculated the redshift of the new augmented event $z_\mathrm{aug}$,
\begin{equation}
\begin{aligned}
z_{\mathrm{aug}}\left(z^{*}\right)=-z^{*}+z_{\mathrm{min}}+z_{\mathrm{max}}\,.
\end{aligned}
\end{equation} 
For the deeper DDF survey, the corresponding $z_{\mathrm{max}}$ was increased by $40\%$, otherwise the same procedure was followed. These limits arise due to the fact that for a given event in the original training set, its GP fit is more reliable close to the observations; hence we limit the GP extrapolation in wavelength when generating synthetic events, which translates into the above redshift constraint. 
This distribution differs slightly from \citet{Boone2019}, which also uses a class-agnostic augmentation. 
We derive the aforementioned redshift limits for augmentation in Appendix \ref{sec:Redshift-events}. 

The process of actually redshifting the light curve after choosing the new redshift is discussed below. 

\subsection{Generating Realistic Synthetic Observations}
\label{subsec:generating_synthetic}

The first step in generating the synthetic light curves is selecting the epochs at which mock observations will be made. Our implementation proceeded as in \citet{Boone2019}; we summarize the approach as follows. First, we stretched the observed epochs of the original event to account for the time dilation due to the difference between the original and augmented redshifts. We also removed any observations that fell outside the observing window as a consequence. Then we randomly picked a target number of observations from a Gaussian mixture model based on the test set\footnote{The model used contained one component with mean $24.5$ and standard deviation $8.5$ for WFD, and two components with probabilities for each component of $[0.34,0.66]$, means of $[57.4,92.8]$ and standard deviations of $[16.5,18.4]$ for DDF. While a mixture model was fitted for both WFD and DDF, a single component was found to be the best fit for WFD.}.
However, this fails to account for the change in the cadence due to redshift augmentation; events shifted to higher redshifts have a lower density of observations than the events observed at those redshifts. In order to account for this, we multiplied this target number by $(1+z_\mathrm{aug})/(1+z_\mathrm{ori})$. We then generated additional observations at the same epochs as existing observations in randomly-selected passbands, associating each synthetic observation with an observed epoch in the original light curve. Further, to avoid creating synthetic light curves where most of the observations are obtained through this procedure, we capped the  number of additional observations generated to be less than $50\%$ of the total number of observations in the original light curve. If this procedure resulted in more observations than the original target number drawn from the Gaussian mixture model, we then randomly dropped observations (original or new) until the target number was reached. Otherwise, to introduce additional variability, we randomly dropped $10\%$ of the synthetic observations.

Once we determined the epochs at which new observations would be generated, we redshifted the light curve as follows. We first computed the central wavelengths of the {\it ugrizy} passbands of the synthetic event as seen at the redshift of the original event. Then, we computed the mean and uncertainty of the GP fit to the original event, at the observed epochs of the original event associated with the synthetic observations but at the redshifted wavelengths. The steps so far dealt with the time dilation but not with the cosmological dimming of the synthetic event. Assuming a standard cosmological model, we redshifted the flux of the synthetic event and its uncertainty, such that it is observed at $z_\mathrm{aug}$. Further details of this redshifting implementation are given in Appendix \ref{sec:Redshift-events}. 

Following \citet{Boone2019}, we then combined the flux uncertainty of the augmented events predicted by the GP in quadrature with a value drawn from the flux uncertainty distribution of the
test set, in order to achieve a more representative flux uncertainty distribution for the augmented training set.
We also drew noise to add to the flux of the augmented events from a Gaussian with standard deviation of the aforementioned value from the flux uncertainty distribution.

Finally, we imposed quality cuts on the synthetic events in order to decide whether to add them to the augmented training set. To make the synthetic events as similar as possible to the test set events, we use a 2-detection trigger based on \plasticc{} \citep{Kessler2019}.
\citet{Boone2019} fitted an error function to the observations from the full dataset to predict the probability of detection as a function of signal-to-noise ratio (S/N), and applied this probabilistic threshold to all observations. \citet{Boone2019} then accepted an event if at least two of its observations were predicted as detected. However, we find that this was insufficient to constraining a GP, thus we required an additional observation, without requiring it to be predicted as detected by the chosen probabilistic detection threshold. Note that all synthetic light curves in the augmented DDF training set meet this quality cut as they are generated with a higher number of observations.

\subsection{Photometric Redshift}
\label{subsec:photometric_redshift}

In order to simulate realistic photometric redshifts for the synthetic events, following \citet{Boone2019} we chose a random event from the $\sim4\%$ of test set events that had a spectroscopic redshift measurement, and calculated the difference between its spectroscopic and photometric redshifts. We then added this difference to the true redshift of the augmented event to generate a photometric redshift.

\subsection{Computational Resources}

We performed our computations on an Intel(R) Xeon(R) CPU
E5-2697 v2 (2.70GHz). Using a single core, the pipeline takes $\sim1$ min
to fit GPs to $1000$ events, and to perform their
wavelet decomposition. Generating a balanced augmented training set
with $4\times10^{4}$ events takes $\sim9$ hrs. Reducing the dimensionality
using PCA takes $\sim30$ min for an augmented training set of $4\times10^{4}$
events and optimizing the LightGBM classifier on the same training
set takes $\sim8$~hrs. After we computed the test set features, generating
predictions with the trained classifier takes $\sim10$ min. Overall,
the entire classification pipeline takes $\sim70$ core hours of computing
time for WFD and $12$ for DDF in this setting.


\begin{figure}
\begin{centering}
\includegraphics[width=0.35\textwidth]{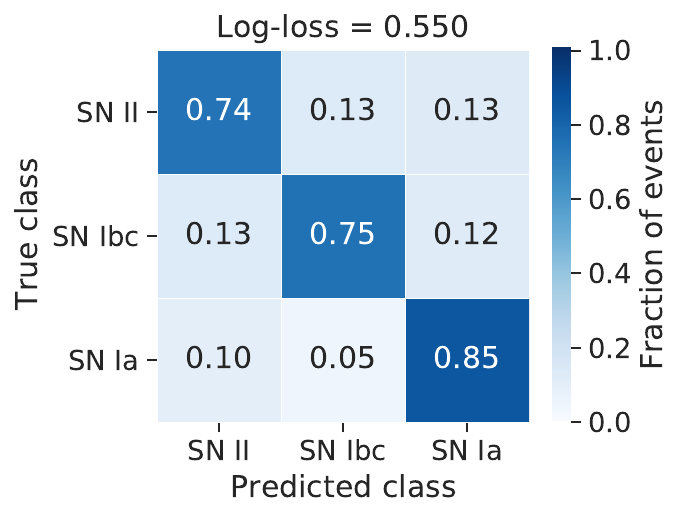}
\par\end{centering}
\caption{WFD test set normalized confusion matrix for the classifier trained on
the augmented training set and its log loss performance. Each event
is assigned to the class with the highest prediction. \label{fig:cm-test}}
\end{figure}

\section{Results and Implications for Observing Strategy \label{sec:results}}

\begin{figure*}
\begin{centering}
\includegraphics[width=0.35\textwidth]{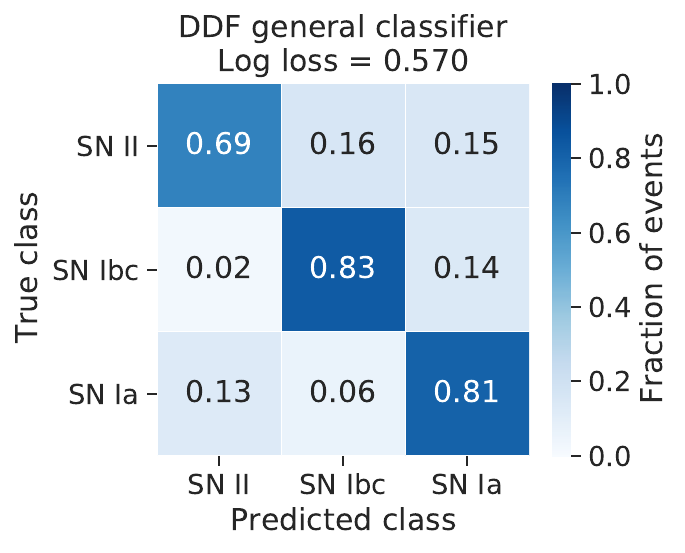}\includegraphics[width=0.35\textwidth]{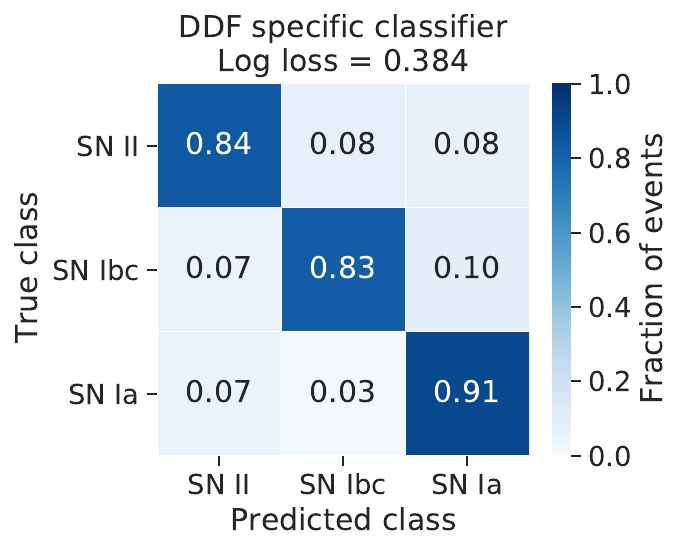}
\par\end{centering}
\caption{DDF test set normalized confusion matrix for the classifier trained
with (left panel) the general WFD+DDF augmented training set and (right panel) with the DDF-only augmented training set. The results show the importance of using an augmented training set customized for the specific survey mode characteristics. \label{fig:cm-ddf}}
\end{figure*}

We now turn to our results on the \plasticc{} dataset and consider in detail their implications for various aspects of the LSST observing strategy. We study classification performance for SNe with different properties within the single simulated observing strategy that is available in \plasticc. We present results related to classification performance for the two different survey modes (WFD and DDF) in Section \ref{subsec:Overall-performance}. We then explore the performance as a function of light curve
length (Section \ref{subsec:Light-curve-length}), median inter-night
gap (Section \ref{subsec:Inter-night-Gaps}), number of gaps $>10$
days (Section \ref{subsec:Inter-night-Gaps}), and number of observations
near the peak (Section \ref{subsec:Observations-Near-Peak}).

\subsection{Survey Mode-specific Augmentation and its Effect on Performance\label{subsec:Overall-performance}}

\begin{figure*}
\begin{centering}
\includegraphics[width=0.3\textwidth]{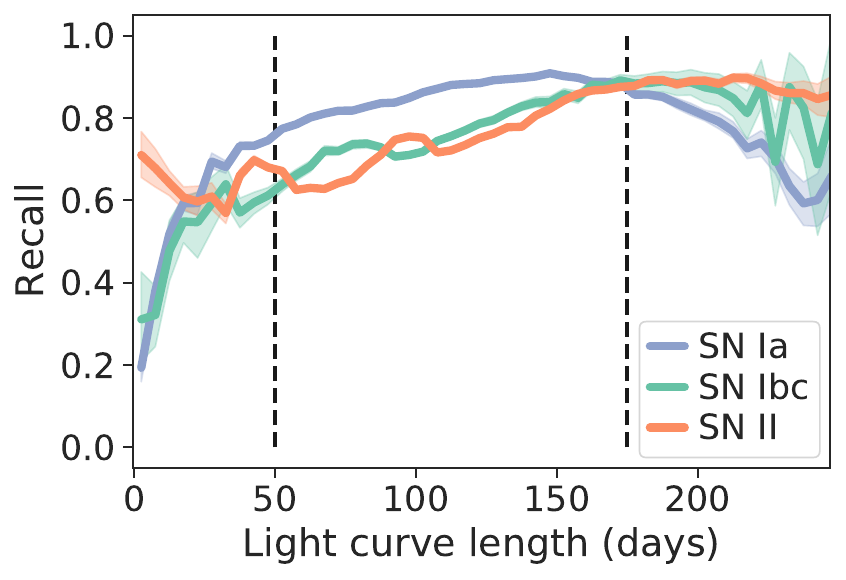}\includegraphics[width=0.3\textwidth]{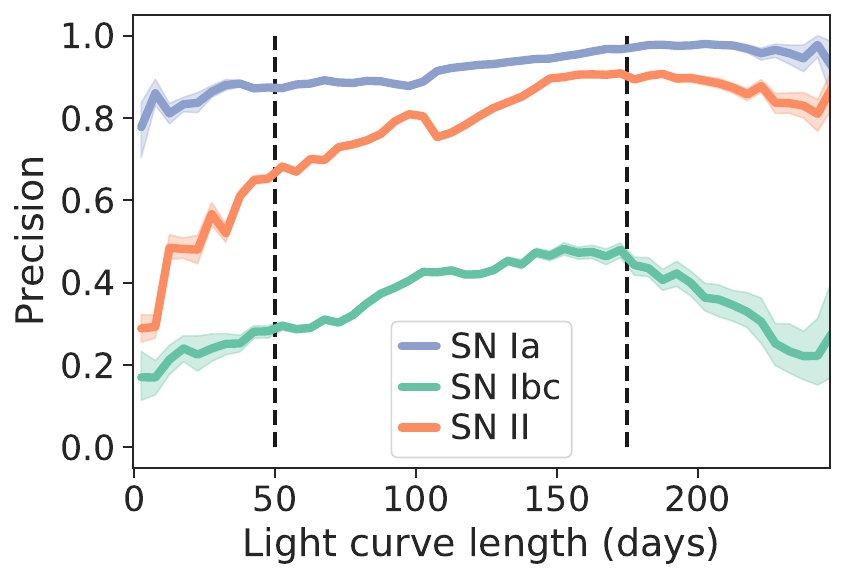}\includegraphics[width=0.3\textwidth]{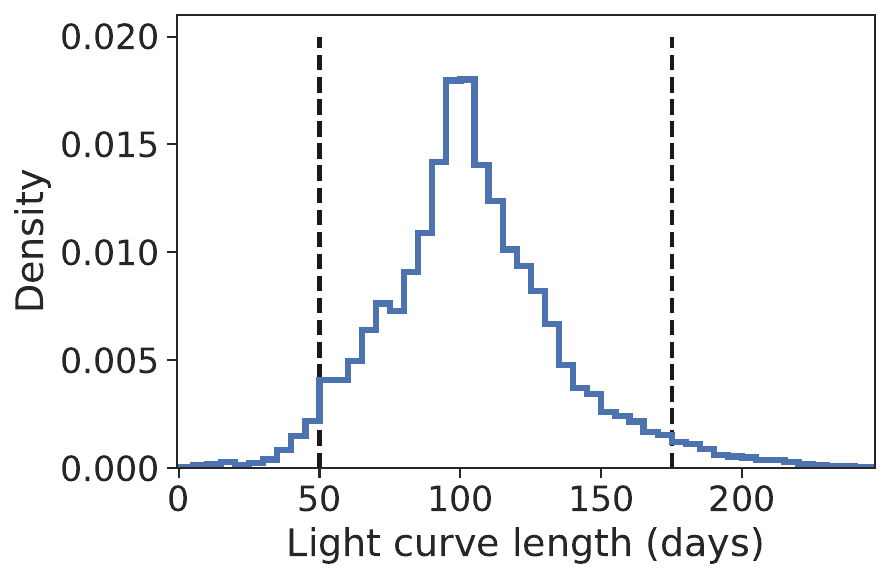}
\par\end{centering}
\caption{WFD test set recall (left panel) and precision (middle panel) as a
function of light curve length per SNe class. The right panel shows the density of events as a function of light curve
length. Because of the low number of events in the tails of the distribution, we restrict our analysis to between $50$ and $175$ days (comprising $94\%$ of the events). Recall and precision increase for longer light curves.\label{fig:lc-length}}
\end{figure*}

Figure \ref{fig:cm-test} shows the confusion matrix for the classifier trained on an augmented WFD training set as described
in Section \ref{subsec:Augmentation}. Despite the use of general wavelet features which were not specifically designed for SNe classfication, the classifier obtains a log-loss of $0.55$. This performance is comparable to that obtained by the top three submissions to
PLAsTiCC for these SN classes \citep{Boone2019,ResultsPlasticc2020}. We note that similar to other classifiers, the performance is weakest for SN Ibc ($75\%$ recall but $39\%$ precision).

The DDF survey contains fainter events with higher cadence, as well as lower flux uncertainty compared to the WFD survey. Unlike the \plasticc{} submissions, we therefore carried out a separate augmentation for this survey mode and built a custom classifier for it, as discussed in Section \ref{subsec:Augmentation}. 

We now compare the DDF test set classification performance when using a classifier which is based on the augmented \plasticc{} training set (which mixes WFD and DDF events) versus one trained on an augmented DDF-only training set. Figure \ref{fig:cm-ddf} shows that 
the classifier optimized for the WFD test set obtains a worse performance on the DDF test set, with a higher log-loss ($0.570$ vs $0.384$) and a lower recall for SNe II and SNe Ia. These results illustrate the vital need for matching augmented training sets to the characteristics of the different survey modes. It also strongly highlights the better classification performance that can be obtained for SNe in the DDF survey compared to the WFD survey.


\subsection{Light Curve Length\label{subsec:Light-curve-length}}

The season length is an important factor for observing strategy, which can be tuned by taking additional observations in suboptimal conditions (such as at high airmass). We compared the classification performance of light curves of different lengths, as a proxy for season length. The right panel of Figure \ref{fig:lc-length} shows that $94\%$ of events in the test set have light curve lengths between $50$--$175$ days; we focus on this interval in the recall (left panel) and precision (middle panel) plots, as outside the range the results are dominated by small-number effects. As expected, events observed for longer are better characterized by the feature extraction step, and hence yield higher recall and precision. Again, we note that for a fixed total exposure time, a reduced season length could be compensated by a higher cadence. Our findings support the minimum five-month season length recommendation in \citet{DESCObservingStrategy2018,osmetrics}.

\begin{figure*}
\begin{centering}
\includegraphics[width=0.3\textwidth]{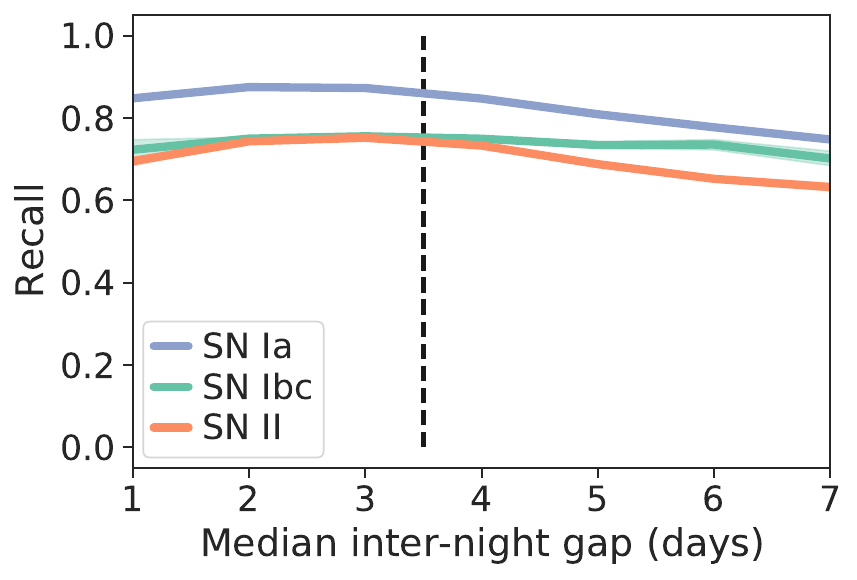}\includegraphics[width=0.3\textwidth]{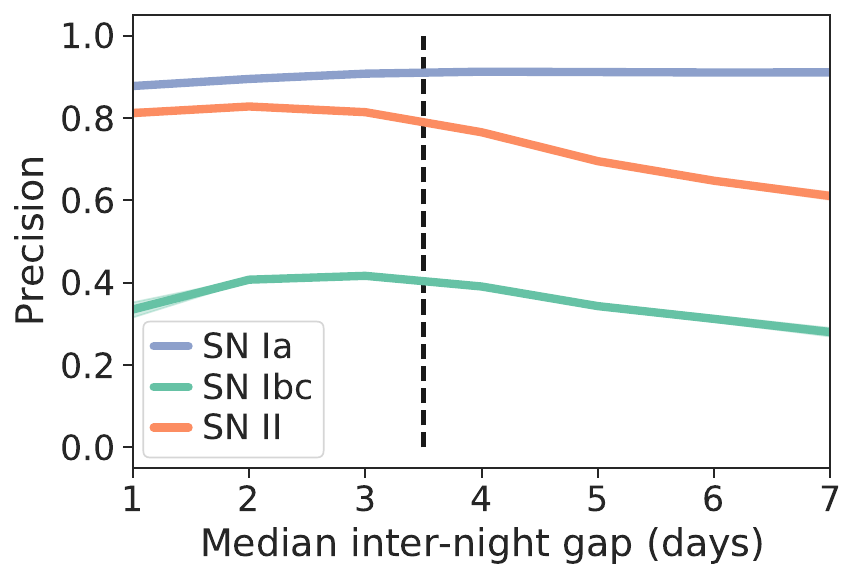}\includegraphics[width=0.3\textwidth]{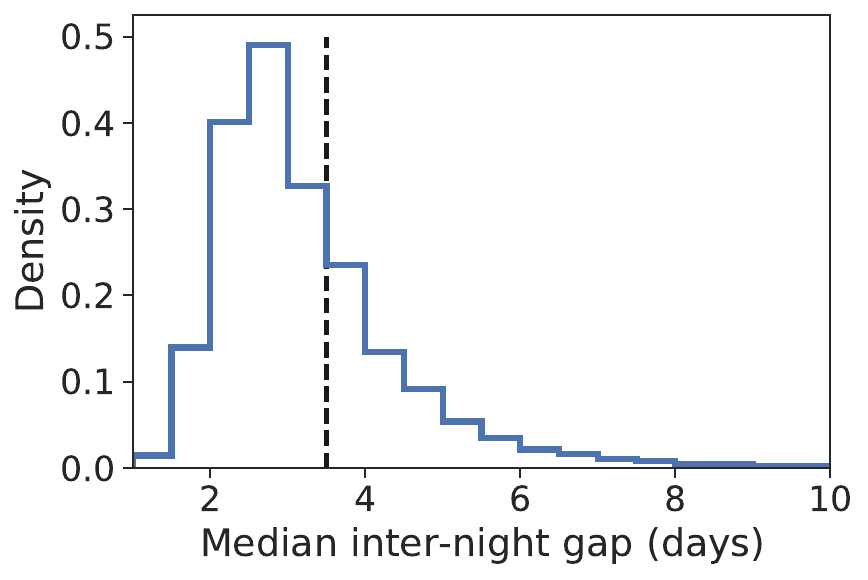}
\par\end{centering}
\caption{WFD test set recall (left panel) and precision (middle panel) as a
function of median inter-night gap per SNe class for events with light
curves between $50$ and $175$ days long. In general, the recall
and precision are higher for events whose median inter-night gap is
$<3.5$ days (left side of the black line). The right panel shows the density of events as a function of median inter-night gap; $\sim64\%$ of SN Ia, $\sim63\%$ of SN Ibc, and $\sim66\%$ of SN II from the test set events have median inter-night gap $<3.5$ days and light curve lengths between $50$ and $175$ days. \label{fig:internight gaps}}
\end{figure*}

\break
\subsection{Inter-night Gaps\label{subsec:Inter-night-Gaps}}

The cadence of observation, as quantified by the inter-night gap when no observations are taken in any passband, is a critical factor in LSST observing strategy that impacts all transient science goals. To investigate this effect, we compared the performance of SNe with different median inter-night gap. The left panel of Figure \ref{fig:internight gaps} shows that
cadence has an important impact on SNe classification; events whose median inter-night gap is $<3.5$ days yield higher recall and precision. Such events comprise nearly $70\%$ of the entire test set. These events are better sampled and thus have a higher light curve quality. Moreover, for a fixed SN Ia recall of $80\%$, the core-collapse SN contamination is $6.8\%$ for events whose median inter-night gap is $<3.5$ days, and $8.0\%$ otherwise.  
These results support previous works such as \citet{DESCObservingStrategy2018,osmetrics} that call for SN Ia light curves to have frequent observations in order to reduce the uncertainty on the cosmological distance modulus. 

\begin{figure*}
\begin{centering}
\includegraphics[width=0.3\textwidth]{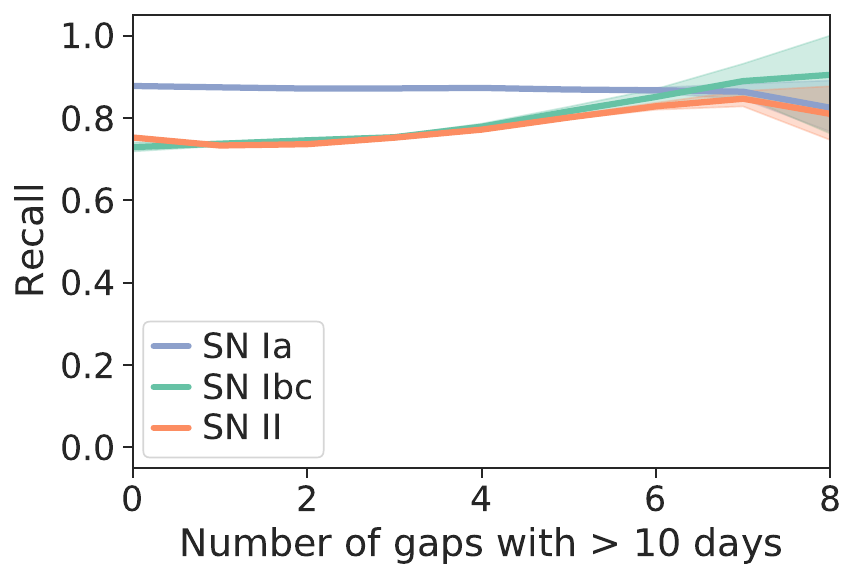}\includegraphics[width=0.3\textwidth]{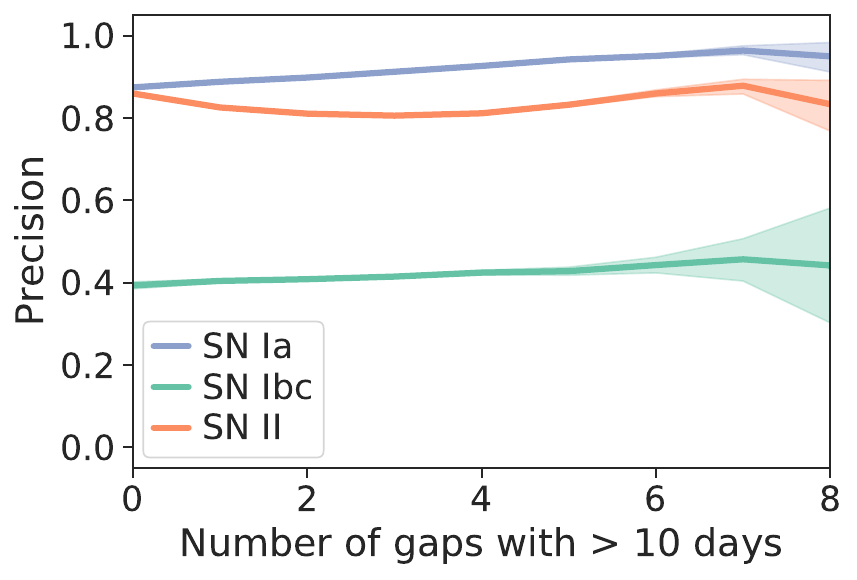}\includegraphics[width=0.3\textwidth]{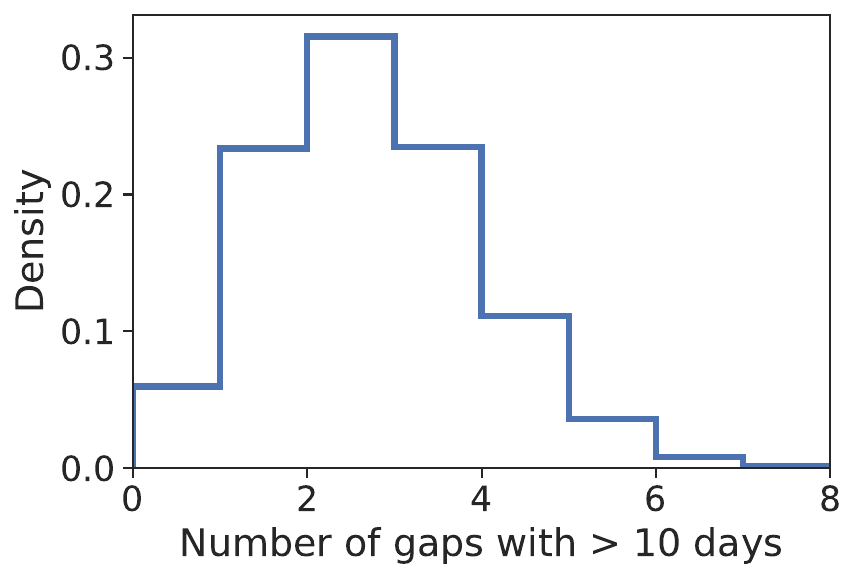}

\includegraphics[width=0.3\textwidth]{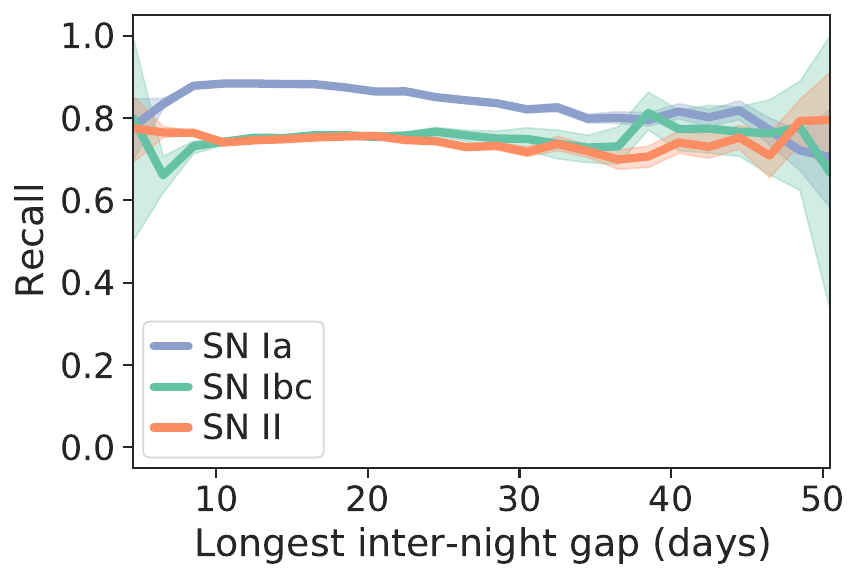}\includegraphics[width=0.3\textwidth]{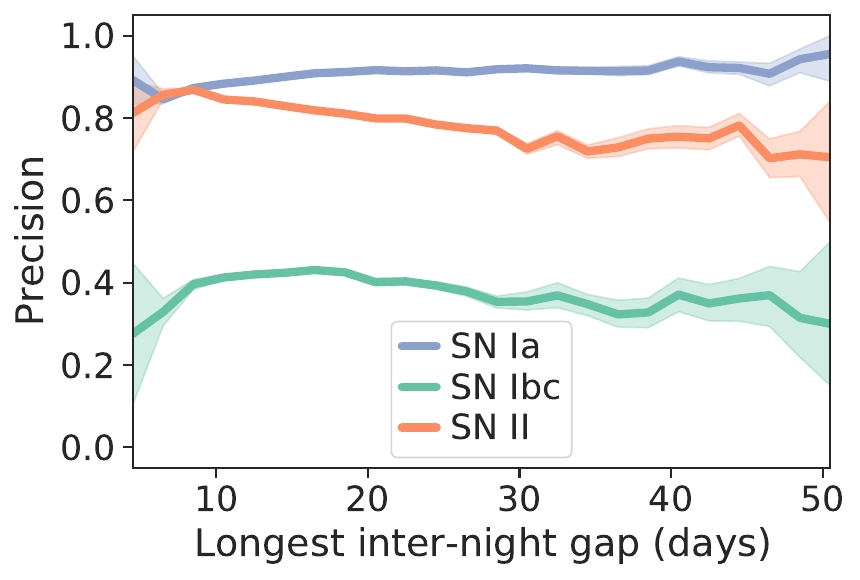}\includegraphics[width=0.3\textwidth]{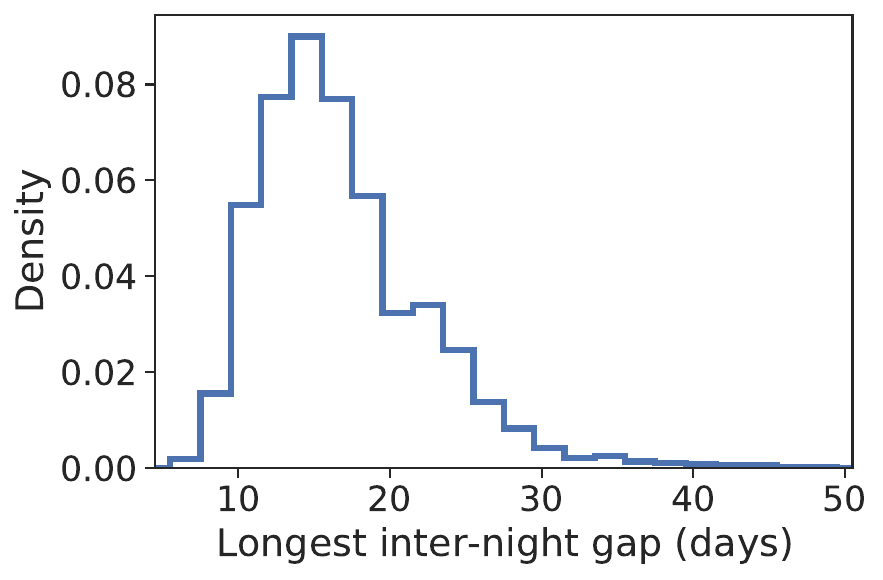}
\par\end{centering}
\caption{WFD test set recall (left panels) and precision (middle panels) 
as a function of the number of gaps  longer than $10$ days (top row) and the length of the longest inter-night gap (bottom row), per SNe class. We
only included events with median inter-night gap $<3.5$ days
and light curves between $50$ and $175$ days long. These results show that large gaps do not significantly impact the SNe classification for this subset. The right panel shows the density of events as a function of the number of large gaps (top row) and the length of the longest inter-night gap (bottom row). We note that the results in the tails of the distribution are dominated by small-number effects.\label{fig:large gaps}}
\end{figure*}

However, the median inter-night gap does not fully capture the impact of gaps in the light curve. A $3.5$-day median inter-night gap does not imply a uniform cadence; it is entirely possible that such light curves contain much larger gaps. To investigate the impact of such `gappy' light curves, we studied the classification performance as a function of the number of large gaps ($>10$ days) in a subsample of events with a median inter-night gap $<3.5$ days.  

The upper left panels of Figure \ref{fig:large gaps} show the recall and precision are broadly independent of the number of large gaps in a light curve\footnote{SN Ibc and SN
II have a small recall increase for higher number of large gaps; we find that these events correspond to longer light curves at lower redshifts, which tend to have a higher recall for SN Ibc and SN II. Note that uncertainties are also larger for cases with a greater number of large gaps due to small-number statistics.}. We tested this with $>20$-day gaps and found similar results. 

\begin{figure*}
\begin{centering}
\includegraphics[width=0.3\textwidth]{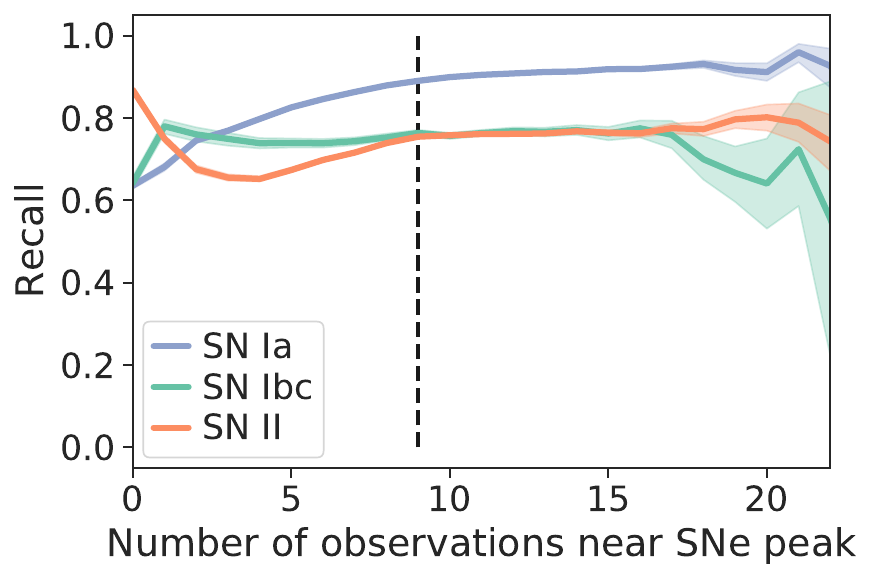}\includegraphics[width=0.3\textwidth]{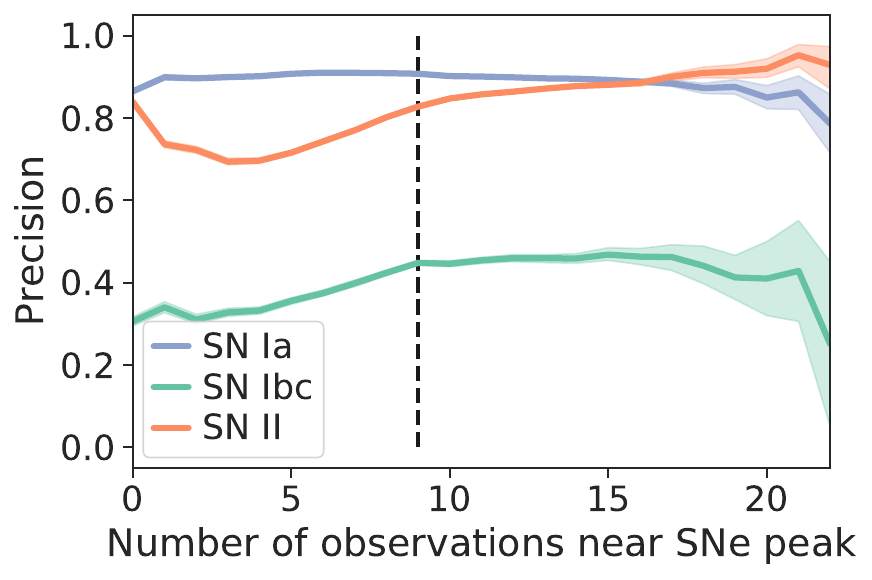}\includegraphics[width=0.3\textwidth]{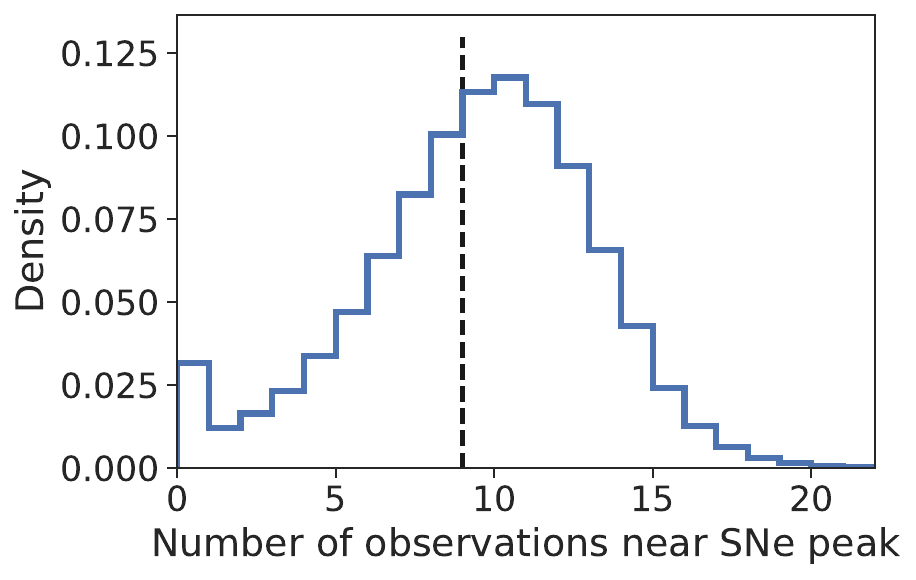}
\par\end{centering}
\caption{WFD test set recall (left panel) and precision (middle panel) as a
function of the number of observation between $10$ days before and
$30$ days after the peak per SNe class. We only included events with
median inter-night gap $<3.5$ days and light curves between
$50$ and $175$ days long. In general, the recall and precision increase
with the number of observations near the peak, until they reach an
approximately constant value for events with $\protect\geq9$ observations. The right panel shows the density of events as a function of number of observations near peak. We note that, as with previous plots, the results in the tails of the distribution are dominated by small-number effects.\label{fig:obs-near-peak}}
\end{figure*}

We expect that the reason for these surprising findings is that the GP fits can still constrain a light curve fit sufficiently well if there are enough points on either side of large gaps. This is demonstrated in Figure \ref{fig:example 2d GP}, which shows an example of a GP fit to an event with four gaps $>10$ days, one of which is $>20$ days. We then compare the classification performance as a function of the length of the longest inter-night gap per light curve, to investigate at which point the performance degrades due to inability of GP fits to constrain a light curve fit. 
The bottom panels of Figure \ref{fig:large gaps} show that the recall and precision of SNe either slowly decrease or remain constant with the increase of the length of longest inter-night gap. 
While previous works recommended a regular cadence without inter-night gaps
larger than $10$~--~$15$ days \citep{DESCObservingStrategy2018,osmetrics}, we find that requiring a median inter-night gap of $<3.5$ days is sufficient for photometric classification methods using GPs that incorporate cross-band information to model the light curves and generate features.

We also find that 98\% of DDF events have a median inter-night gap of $<3.5$ days, and hence the DDF sample performs uniformly well independent of the inter-night gap.


\subsection{Observations Near Peak\label{subsec:Observations-Near-Peak}}
Obtaining observations near the peak of a SN Ia light curve is generally considered critical to obtain a reliable cosmological distance modulus. To investigate whether SNe classification has a similar requirement, we analyzed the classification performance as a function of the number of observations near the peak (defined as $10$ days before and $30$ days after the peak). We estimated the peak time as the moment that maximizes the
GP fit predicted flux. Figure \ref{fig:obs-near-peak}
shows that the recall and precision generally increase with the number of
observations near the peak, reaching a constant value for events
with more than nine observations. This improvement in performance is likely due to better characterization of light curve shape. However, since we cannot predict when a SN will be observed, this result only further demonstrates the importance of frequent observations to increase the likelihood of obtaining observations near the peak. These results agree with \citet{Takahashi2020}, who found that SNe light curves without observations near the peak were more often misclassified.


\section{Discussion and Conclusions} \label{sec:conclusion}

We have presented a quantitative analysis of the impact of various factors related to the LSST observing strategy on the performance of SNe photometric classification, using the \plasticc\ simulation. We use the photometric transient classification library \texttt{snmachine}, based on model-independent wavelet features (instead of specialized features constructed using domain knowledge about SNe). In line with previous studies using the \plasticc{} data, we confirm that augmentation for a number of aspects (the photometric redshift distribution per supernovae class, the distribution of the observing cadence, and the flux uncertainty distribution) is crucial for obtaining a representative training set for machine learning classification. 

Our classifier yields similar performance
to the top \plasticc{} submission \citep{Boone2019,ResultsPlasticc2020}
and competitive results in core-collapse SN contamination \citep[see below;][]{Kessler2017,Jones2017},
which is essential for measurements of the dark energy equation of
state parameter. We obtain a core-collapse SN contamination of $8.3\%$ (for SNe predicted to be SN Ia with~$>50\%$ probability) which is comparable to the $\sim5\%$ contamination obtained in \citet{Jones2018} with Pan-STARRS SNe. This could be further improved by optimizing the classifier for SN Ia classification rather than overall classification performance as was done in \plasticc. \citet{Jones2018} demonstrated that this level of contamination provides competitive cosmological constraints when using a Bayesian methodology to marginalize over the contamination. Hence, we expect our contamination levels to also be acceptable for cosmology when used along with a Bayesian methodology such as Bayesian Estimation Applied to Multiple Species \citep{Kunz2007,Lochner2013,Roberts2017,Jones2018}.

Turning to the question of how observing strategy impacts classification, our results demonstrate the importance of  customized training set augmentation for each LSST survey mode (WFD and DDF). We find that the season length is important -- in general, better classification performance is obtained for longer light curves. This supports the minimum five-month season length recommendation in \citet{DESCObservingStrategy2018, osmetrics}. Further, we show that good classification performance requires a cadence with a median inter-night gap of $<3.5$ days. Surprisingly, however, we find that large gaps of~$>10$ days do not impact the classification performance for events exhibiting such a cadence, due to the ability of the Gaussian process methods we use to interpolate such gaps effectively. Finally, a regular cadence which achieves $>9$ observations near the peak of the light curve provides effective classification performance. In Appendix \ref{sec:compare-avocado} we show that these results also hold if we replace our classification predictions with the predictions obtained by \citet{Boone2019}, who use a different feature set and an independent classification framework with somewhat different augmentation choices.

These results provide guidance for further refinement of the LSST observing strategy on the question of SNe photometric classification. While the \plasticc{} simulation used in this analysis has an outdated cadence, we expect our general conclusions to hold for any reasonable variation currently under consideration. Our augmentation and classification pipeline will be used in the future to study the SNe classification performance of more recent observing strategy simulations in detail. 

Since the release of \plasticc, new and more realistic observing strategy simulations have been released. These simulations include improvements to the scheduler, more realistic weather, and changes to the cadence in different bands. While new transient simulations using the more recent baseline observing strategy may result in different classification performance, we still expect our broad conclusions to remain unchanged. Future work will include investigating the dependence of classification performance on different observing strategy simulations.

With this paper we publicly release the photometric transient classification library  \texttt{snmachine}\footnote{\url{https://github.com/LSSTDESC/snmachine}}. The library also contains some example Jupyter notebooks which can be used to reproduce this work. In the future, the \texttt{snmachine} pipeline will be extended to facilitate the classification of other transient classes.


\acknowledgments


This paper has undergone internal review in the LSST Dark Energy Science Collaboration. The authors would like to thank Kyle Boone, Patrick D. Aleo, and Philippe Gris for their helpful comments and reviews.

$\,$


Author contributions are listed below.

CSA: software, validation, formal analysis, investigation, data curation, writing (original draft), visualization

HVP: conceptualization, methodology, validation and interpretation, supervision, writing (original draft; review \& editing), funding acquisition

ML: conceptualization, methodology, software, validation and interpretation, writing (original draft; review \& editing)

JDM: conceptualization, methodology, validation and interpretation, supervision, writing (reviewing \& editing)

TA: software, data curation, writing (editing)

RB: software 

$\,$

We thank Christian N. Setzer for useful discussions and for setting up the initial code framework to handle the \plasticc\ light curves and metadata. We also thank Gautham Narayan for useful discussions and for contributing to reviewing the updated \texttt{snmachine} software.
This work was partially enabled by funding from the UCL Cosmoparticle Initiative.
This work used facilities provided by the UCL Cosmoparticle Initiative; and we thank the HPC systems manager Edd Edmondson for his support.
HVP and RB were partially supported by the research environment grant ``Gravitational Radiation and Electromagnetic Astrophysical Transients (GREAT)" funded by the Swedish Research council (VR) under Dnr 2016-06012, and the research project grant ``Understanding the Dynamic Universe" funded by the Knut and Alice Wallenberg Foundation under Dnr KAW 2018.0067.
ML acknowledges support from South African Radio Astronomy Observatory and the National Research Foundation (NRF) towards this research. Opinions expressed and conclusions arrived at, are those of the authors and are not necessarily to be attributed to the NRF.
TA is supported in part by STFC.

The DESC acknowledges ongoing support from the Institut National de 
Physique Nucl\'eaire et de Physique des Particules in France; the 
Science \& Technology Facilities Council in the United Kingdom; and the
Department of Energy, the National Science Foundation, and the LSST 
Corporation in the United States.  DESC uses resources of the IN2P3 
Computing Center (CC-IN2P3--Lyon/Villeurbanne - France) funded by the 
Centre National de la Recherche Scientifique; the National Energy 
Research Scientific Computing Center, a DOE Office of Science User 
Facility supported by the Office of Science of the U.S.\ Department of
Energy under Contract No.\ DE-AC02-05CH11231; STFC DiRAC HPC Facilities, 
funded by UK BIS National E-infrastructure capital grants; and the UK 
particle physics grid, supported by the GridPP Collaboration.  This 
work was performed in part under DOE Contract DE-AC02-76SF00515.




\software{Astropy \citep{astropy2013,astropy2018}, George \citep{george2014},
Jupyter \citep{jupyter2016}, LightGBM \citep{lgbm2016,ke2017lightgbm,lgbm2017},
Matplotlib \citep{matplotlib2007,matplotlib2020}, NumPy \citep{harris2020array},
pandas \citep{mckinney2010pandas,reback2020pandas}, pickle \citep{van1995python},
pytest \citep{pytest6.2.2}, pywt \citep{Lee2019,Lee2019a}, scikit-learn
\citep{scikit-learn2011}, SciPy \citep{2020SciPy}, seaborn \citep{Waskom2020seabor},
snmachine \citep{Lochner2016}}

\appendix
\section{Redshifting Implementation for Augmentation\label{sec:Redshift-events}}

In this appendix we provide further details of the augmentation procedure
described in Section \ref{subsec:Augmentation}. In particular, we
present the technique for redshifting a light curve, and derive the redshift limits for augmentation shown in Equation~(\ref{eq:redshift range z-max-scale mini}).

Consider a multi-band SN light curve at redshift $z_{\mathrm{ori}}$, from which we want to create a synthetic multi-band light curve at redshift $z_{\mathrm{aug}}$. For each epoch, the spectrum of the new synthetic SN is
\begin{equation}
f_{\lambda;\,\mathrm{aug}}(\lambda)=\dfrac{1+z_{\mathrm{ori}}}{1+z_{\mathrm{aug}}}\left[\dfrac{d_{\mathrm{L}}\left(z_{\mathrm{ori}}\right)}{d_{\mathrm{L}}\left(z_{\mathrm{aug}}\right)}\right]^{2}f_{\lambda}\left(\dfrac{1+z_{\mathrm{ori}}}{1+z_{\mathrm{aug}}}\lambda\right)\,,\,\label{eq:spectrum expression}
\end{equation}
where $\lambda$ is the observed wavelength, $d_{\mathrm{L}}$ is
the luminosity distance, and $f_{\lambda}$ is the spectrum of the
original event. Note that the spectrum of the synthetic SN depends
on the original spectrum evaluated at redshifted wavelengths. The two-dimensional GP fit described in Section \ref{subsec:Model-light-curves} then models
the convolution of the original spectrum with the {\it ugrizy} passbands to predict the measured flux. Thus, for each epoch of the synthetic SN,
we estimated the flux in the original event at each redshifted passband $\mathrm{b}$ (where $\mathrm{b}=u,g,r,i,z,y$) as 
\begin{equation}
F_{\mathrm{ori\,b}}=\mathcal{GP}\left(\dfrac{1+z_{\mathrm{ori}}}{1+z_{\mathrm{aug}}}\lambda_{\mathrm{b}}\right)\,,\label{eq:gp-pred}
\end{equation}
where $\mathcal{GP}$ represents the mean of the GP fit used to model
the flux observations of the original SNe, and $\lambda_{\mathrm{b}}$
is the central wavelength of passband $\mathrm{b}$. We calculated
these central wavelengths using the LSST throughputs\footnote{\href{https://github.com/lsst/throughputs}{https://github.com/lsst/throughputs}}.
Similarly, we estimated the flux uncertainty in each passband as the
uncertainty of the GP fit.

Finally, we adjusted the fluxes of the synthetic event and their uncertainties
to the desired redshift $z_{\mathrm{aug}}$. We assumed a
flat $\Lambda$CDM cosmology with $H_0=70$ km/s/Mpc and
$\Omega_\mathrm{m}=0.3$. We estimated the flux of the synthetic event in each passband $\mathrm{b}$ as 
\begin{equation}
F_{\mathrm{aug\,b}}=\dfrac{1+z_{\mathrm{ori}}}{1+z_{\mathrm{aug}}}\left[\dfrac{d_{\mathrm{L}}\left(z_{\mathrm{ori}}\right)}{d_{\mathrm{L}}\left(z_{\mathrm{aug}}\right)}\right]^{2}F_{\mathrm{ori\,b}}\,,\label{eq:scale flux}
\end{equation}
and estimated its uncertainty similarly.

\begin{figure}
\begin{centering}
\includegraphics[width=0.4\textwidth]{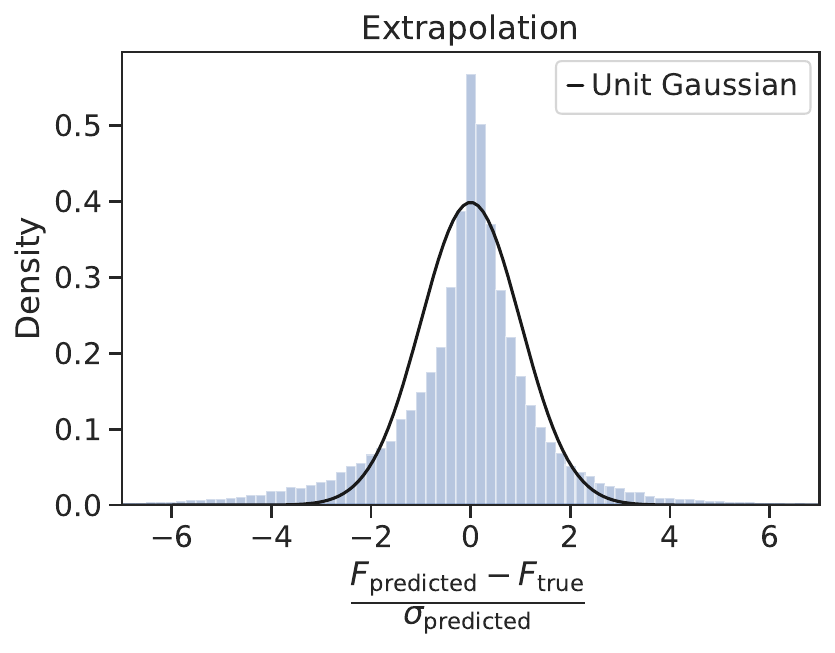}
\par\end{centering}
\caption{
Distribution of the GP errors resulting from extrapolating GP fits
to {\it u} or {\it y} passbands. $F_{\mathrm{true}}$ is the true
flux of an observation, and $F_{\mathrm{predicted}}$ and $\sigma_{\mathrm{predicted}}$
are the flux and its uncertainty predicted by a GP fit at the corresponding
epoch and passband. An ideal error estimation results in a unit Gaussian
(black line).
\label{fig:gp extrapolate}}
\end{figure}

As previously discussed in Section \ref{subsec:redshift_augmentation}, the GP fit is more reliable
close to observations. To test the GP extrapolation, for every SNe
in the training set, we fitted a GP with the observations in the {\it ugriz}
passbands. Then, we compared the observed flux in the {\it y} passband
 with the flux predictions of the GP fit at the same epochs.
Additionally, we repeated this procedure to test the GP extrapolation
in the {\it u} passband using the observations in the {\it grizy}
passbands. Figure \ref{fig:gp extrapolate} shows the GP is reliable
despite underestimating some flux errors. Since the GP errors increase at wavelengths far from the original observation, we restricted our extrapolation
to minimum ($\lambda_{g}-\lambda_{u}$) and maximum ($\lambda_{y}-\lambda_{z}$) wavelength ranges.
Thus, when generating a synthetic SN at higher redshifts, we have that $\ensuremath{\lambda_{u}-(1+z_{\mathrm{ori}})/(1+z_{\mathrm{aug}})}\lambda_{u}\leq\lambda_{g}-\lambda_{u}$. Similarly, for events generated at lower redshifts, we obtain the redshift limits
for augmentation presented in Section~\ref{subsec:redshift_augmentation}: 
\begin{equation}
\begin{aligned}z_{\mathrm{min}} & =\max\left\{ 0,\left(1+z_{\mathrm{ori}}\right)\left(2-\dfrac{\lambda_{\mathrm{z}}}{\lambda_{\mathrm{y}}}\right)^{-1}-1\right\} \\
 & \approx\max\left\{ 0,0.90\,z_{\mathrm{ori}}-0.10\right\}\,, \\
z_{\mathrm{max}} & =\left(1+z_{\mathrm{ori}}\right)\left(2-\dfrac{\lambda_{\mathrm{g}}}{\lambda_{\mathrm{u}}}\right)^{-1}-1\\
 & \approx1.43\,z_{\mathrm{ori}}+0.43 \,.
\end{aligned}
\,.\label{eq:redshift range z-max-scale}
\end{equation}

\section{Comparison with other PLASTICC classifiers\label{sec:compare-avocado}}

\begin{figure}
\begin{centering}
Classification predictions from present work

\includegraphics[width=0.3\textwidth]{recall_lc_length.pdf}\includegraphics[width=0.3\textwidth]{recall_median_gap_small.pdf}\includegraphics[width=0.3\textwidth]{recall_big_gaps_10.pdf}

Classification predictions from \citet{Boone2019}

\includegraphics[width=0.3\textwidth]{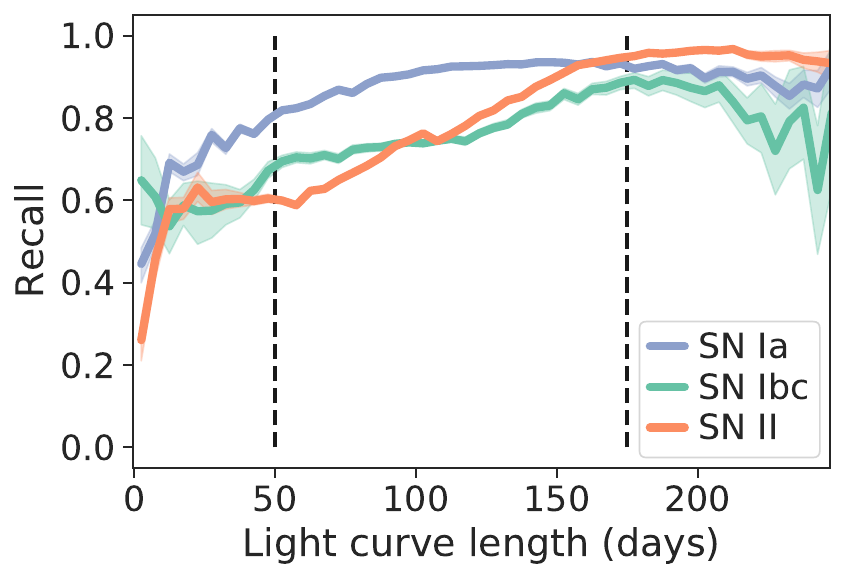}\includegraphics[width=0.3\textwidth]{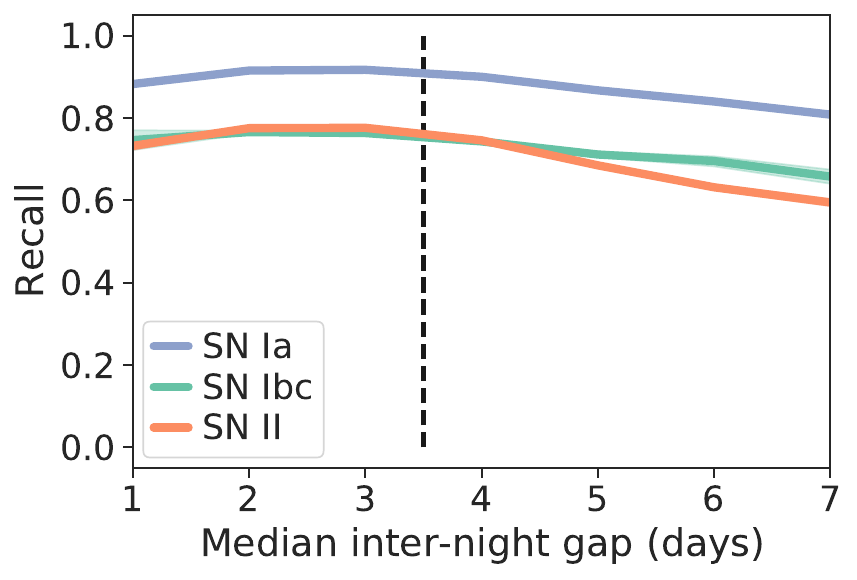}\includegraphics[width=0.3\textwidth]{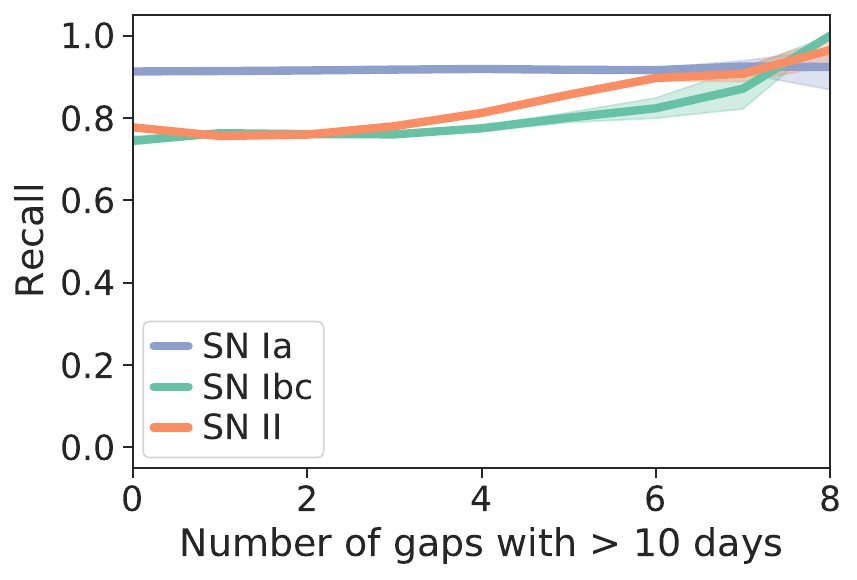}

Classification predictions from present work

\includegraphics[width=0.3\textwidth]{precision_lc_length.pdf}\includegraphics[width=0.3\textwidth]{precision_median_gap_small.pdf}\includegraphics[width=0.3\textwidth]{precision_big_gaps_10.pdf}

Classification predictions from \citet{Boone2019}

\includegraphics[width=0.3\textwidth]{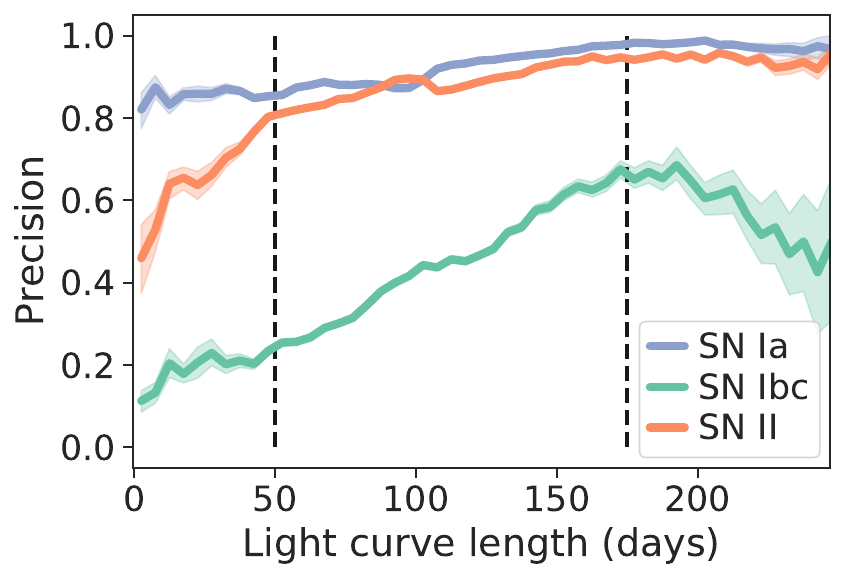}\includegraphics[width=0.3\textwidth]{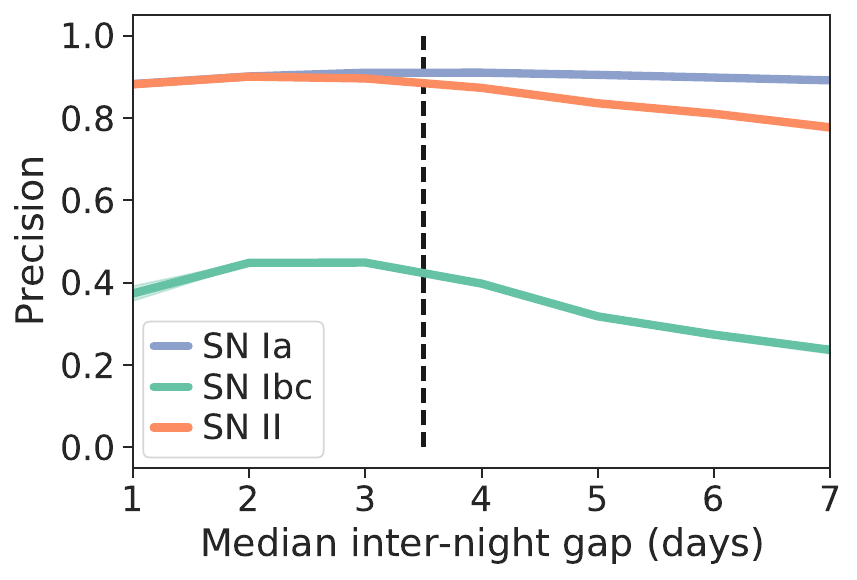}\includegraphics[width=0.3\textwidth]{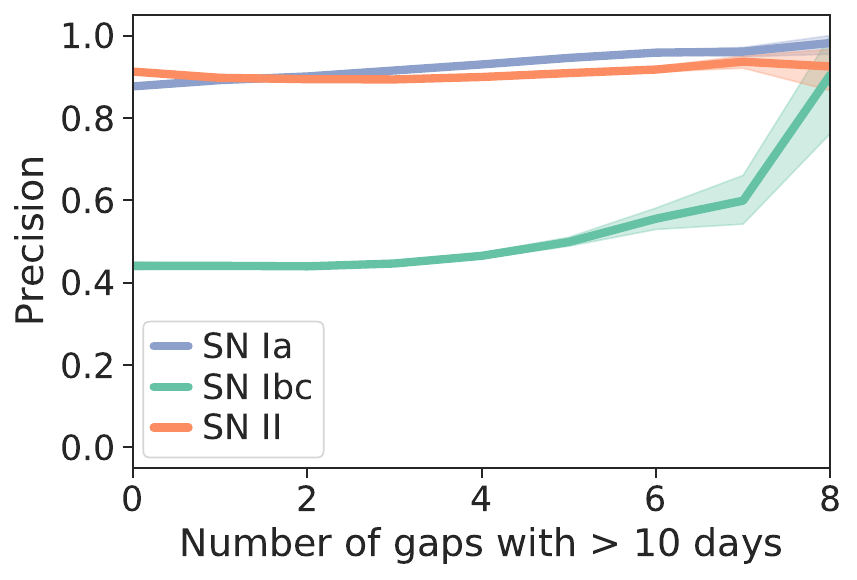}

\end{centering}
\caption{WFD test set recall (top two rows) and precision (bottom two rows)
as a function of light curve length (left panels), median inter-night
gap (middle panels), and number of gaps larger than $10$ days (right
panels), per SNe class. In the first and third rows we reproduce the
results of this work previously shown in Figures \ref{fig:lc-length}, \ref{fig:internight gaps}, and \ref{fig:large gaps}. In the second
and fourth rows we show the classification predictions obtained by
\citet{Boone2019}. \label{fig:avo-comp-1}}
\end{figure}

\begin{figure}
\begin{centering}
Classification predictions from present work

\includegraphics[width=0.3\textwidth]{recall_longest_gap.pdf}\includegraphics[width=0.3\textwidth]{recall_near_peak_nobs.pdf}

Classification predictions from \citet{Boone2019}

\includegraphics[width=0.3\textwidth]{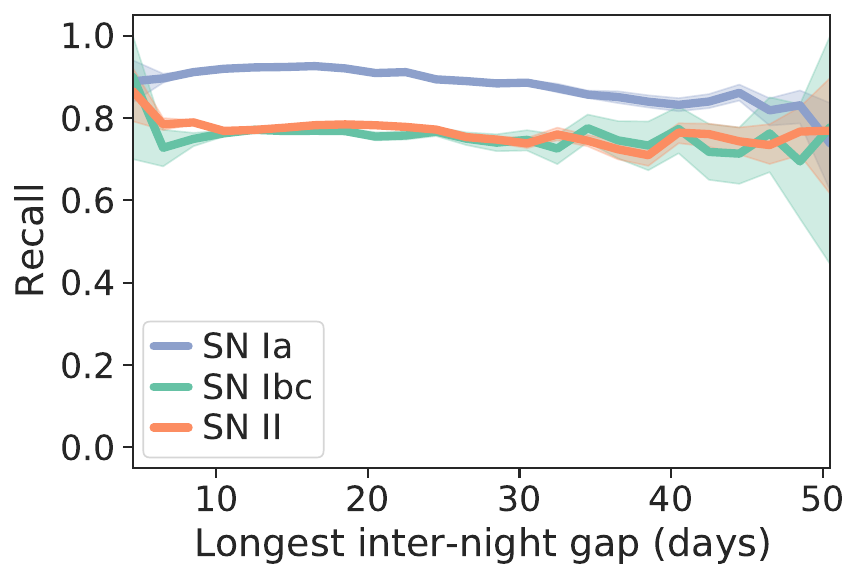}\includegraphics[width=0.3\textwidth]{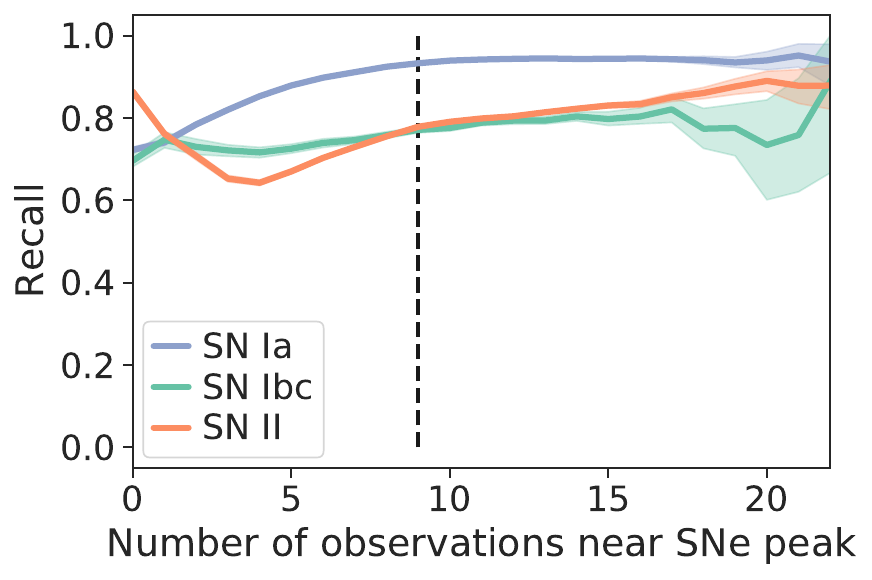}

Classification predictions from present work

\includegraphics[width=0.3\textwidth]{precision_longest_gap.pdf}\includegraphics[width=0.3\textwidth]{precision_near_peak_nobs.pdf}

Classification predictions from \citet{Boone2019}

\includegraphics[width=0.3\textwidth]{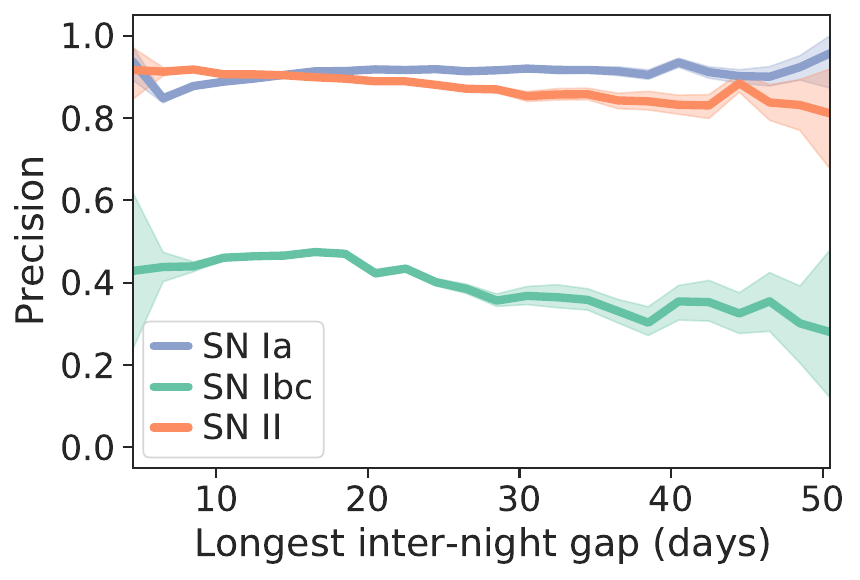}\includegraphics[width=0.3\textwidth]{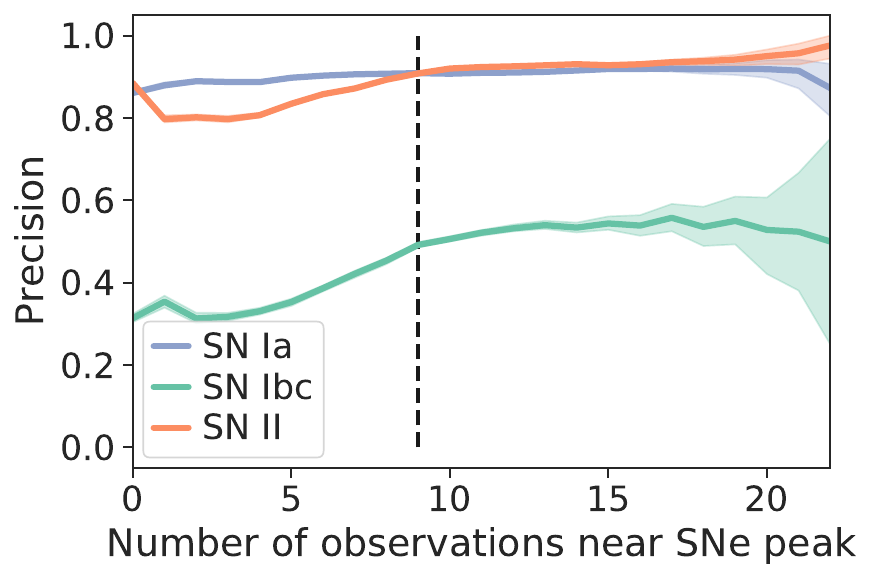}

\end{centering}
\caption{WFD test set recall (top two rows) and precision (bottom two rows)
as a function of the length of the longest inter-night gap (left panels) and the number of observations between $10$ days before and $30$ days after the peak length (right
panels), per SNe class. In the first and third rows we reproduce the
results of this work previously shown in Figures \ref{fig:large gaps} and \ref{fig:obs-near-peak}. In the second
and fourth rows we show the classification predictions obtained by
\citet{Boone2019}.\label{fig:avo-comp-2}}
\end{figure}

Section \ref{sec:results} presented the results of our classifier on the impact of observing strategy on photometric classification. In this appendix,
we show that our results are generalizable beyond our classification pipeline,
by replacing the our classification predictions with those obtained
by \citet{Boone2019}. We use the publicly available predictions
for SN Ia, SN Ibc, and SN II in the test set\footnote{\url{http://supernova.lbl.gov/avocado_plasticc/predictions/predictions_plasticc_test_flat_weight.csv}};
we choose the predictions obtained with a classifier optimized on
the log-loss metric, which equally weights all the \plasticc{} classes
($w_{i}=1$ in Equation \ref{eq:logloss plasticc}). The choice of this flat-weighted metric reduces the impact of additional classes upweighted in the original challenge, but unused in the present work. Figures
\ref{fig:avo-comp-1} and \ref{fig:avo-comp-2} show that the classifier used in \citet{Boone2019} has the same performance behavior as ours. This further indicates that our conclusions are general and not an artifact of our classification architecture.

\clearpage
\bibliography{LiteratureReview}

\begin{thebibliography}{}
\expandafter\ifx\csname natexlab\endcsname\relax\def\natexlab#1{#1}\fi
\providecommand{\url}[1]{\href{#1}{#1}}
\providecommand{\dodoi}[1]{doi:~\href{http://doi.org/#1}{\nolinkurl{#1}}}
\providecommand{\doeprint}[1]{\href{http://ascl.net/#1}{\nolinkurl{http://ascl.net/#1}}}
\providecommand{\doarXiv}[1]{\href{https://arxiv.org/abs/#1}{\nolinkurl{https://arxiv.org/abs/#1}}}

\bibitem[{{Ambikasaran} {et~al.}(2014){Ambikasaran}, {Foreman-Mackey},
  {Greengard}, {Hogg}, \& {O'Neil}}]{george2014}
{Ambikasaran}, S., {Foreman-Mackey}, D., {Greengard}, L., {Hogg}, D.~W., \&
  {O'Neil}, M. 2014, arXiv preprint arXiv:1403.6015v2.
\newblock \url{http://arxiv.org/abs/1403.6015}

\bibitem[{Astier {et~al.}(2006)Astier, Guy, Regnault, Pain, Aubourg, Balam,
  Basa, Carlberg, Fabbro, Fouchez, Hook, Howell, Lafoux, Neill,
  Palanque-Delabrouille, Perrett, Pritchet, Rich, Sullivan, Taillet, Aldering,
  Antilogus, Arsenijevic, Balland, Baumont, Bronder, Courtois, Ellis, Filiol,
  Gon{\c{c}}alves, Goobar, Guide, Hardin, Lusset, Lidman, McMahon, Mouchet,
  Mourao, Perlmutter, Ripoche, Tao, \& Walton}]{Astier2006}
Astier, P., Guy, J., Regnault, N., {et~al.} 2006, Astronomy {\&} Astrophysics,
  447, 31, \dodoi{10.1051/0004-6361:20054185}

\bibitem[{{Astropy Collaboration} {et~al.}(2013){Astropy Collaboration},
  {Robitaille}, {Tollerud}, {Greenfield}, {Droettboom}, {Bray}, {Aldcroft},
  {Davis}, {Ginsburg}, {Price-Whelan}, {Kerzendorf}, {Conley}, {Crighton},
  {Barbary}, {Muna}, {Ferguson}, {Grollier}, {Parikh}, {Nair}, {Unther},
  {Deil}, {Woillez}, {Conseil}, {Kramer}, {Turner}, {Singer}, {Fox}, {Weaver},
  {Zabalza}, {Edwards}, {Azalee Bostroem}, {Burke}, {Casey}, {Crawford},
  {Dencheva}, {Ely}, {Jenness}, {Labrie}, {Lim}, {Pierfederici}, {Pontzen},
  {Ptak}, {Refsdal}, {Servillat}, \& {Streicher}}]{astropy2013}
{Astropy Collaboration}, {Robitaille}, T.~P., {Tollerud}, E.~J., {et~al.} 2013,
  Astronomy \& Astrophysics, 558, A33, \dodoi{10.1051/0004-6361/201322068}

\bibitem[{{Astropy Collaboration} {et~al.}(2018){Astropy Collaboration},
  {Price-Whelan}, {Sip{\H{o}}cz}, {G{\"u}nther}, {Lim}, {Crawford}, {Conseil},
  {Shupe}, {Craig}, {Dencheva}, {Ginsburg}, {Vand erPlas}, {Bradley},
  {P{\'e}rez-Su{\'a}rez}, {de Val-Borro}, {Aldcroft}, {Cruz}, {Robitaille},
  {Tollerud}, {Ardelean}, {Babej}, {Bach}, {Bachetti}, {Bakanov}, {Bamford},
  {Barentsen}, {Barmby}, {Baumbach}, {Berry}, {Biscani}, {Boquien}, {Bostroem},
  {Bouma}, {Brammer}, {Bray}, {Breytenbach}, {Buddelmeijer}, {Burke},
  {Calderone}, {Cano Rodr{\'\i}guez}, {Cara}, {Cardoso}, {Cheedella}, {Copin},
  {Corrales}, {Crichton}, {D'Avella}, {Deil}, {Depagne}, {Dietrich}, {Donath},
  {Droettboom}, {Earl}, {Erben}, {Fabbro}, {Ferreira}, {Finethy}, {Fox},
  {Garrison}, {Gibbons}, {Goldstein}, {Gommers}, {Greco}, {Greenfield},
  {Groener}, {Grollier}, {Hagen}, {Hirst}, {Homeier}, {Horton}, {Hosseinzadeh},
  {Hu}, {Hunkeler}, {Ivezi{\'c}}, {Jain}, {Jenness}, {Kanarek}, {Kendrew},
  {Kern}, {Kerzendorf}, {Khvalko}, {King}, {Kirkby}, {Kulkarni}, {Kumar},
  {Lee}, {Lenz}, {Littlefair}, {Ma}, {Macleod}, {Mastropietro}, {McCully},
  {Montagnac}, {Morris}, {Mueller}, {Mumford}, {Muna}, {Murphy}, {Nelson},
  {Nguyen}, {Ninan}, {N{\"o}the}, {Ogaz}, {Oh}, {Parejko}, {Parley}, {Pascual},
  {Patil}, {Patil}, {Plunkett}, {Prochaska}, {Rastogi}, {Reddy Janga},
  {Sabater}, {Sakurikar}, {Seifert}, {Sherbert}, {Sherwood-Taylor}, {Shih},
  {Sick}, {Silbiger}, {Singanamalla}, {Singer}, {Sladen}, {Sooley},
  {Sornarajah}, {Streicher}, {Teuben}, {Thomas}, {Tremblay}, {Turner},
  {Terr{\'o}n}, {van Kerkwijk}, {de la Vega}, {Watkins}, {Weaver}, {Whitmore},
  {Woillez}, {Zabalza}, \& {Astropy Contributors}}]{astropy2018}
{Astropy Collaboration}, {Price-Whelan}, A.~M., {Sip{\H{o}}cz}, B.~M., {et~al.}
  2018, The Astronomical Journal, 156, 123, \dodoi{10.3847/1538-3881/aabc4f}

\bibitem[{Barbier {et~al.}(2016)Barbier, Dia, Macris, Krzakala, Lesieur, \&
  Zdeborov\'{a}}]{lgbm2016}
Barbier, J., Dia, M., Macris, N., {et~al.} 2016, in Advances in Neural
  Information Processing Systems, ed. D.~Lee, M.~Sugiyama, U.~Luxburg,
  I.~Guyon, \& R.~Garnett, Vol.~29 (Curran Associates, Inc.).
\newblock
  \url{https://proceedings.neurips.cc/paper/2016/file/621bf66ddb7c962aa0d22ac97d69b793-Paper.pdf}

\bibitem[{Boone(2019)}]{Boone2019}
Boone, K. 2019, The Astronomical Journal, 158, 257,
  \dodoi{10.3847/1538-3881/ab5182}

\bibitem[{Carrick {et~al.}(2021)Carrick, Hook, Swann, Boone, Frohmaier, Kim, \&
  Sullivan}]{Carrick2021}
Carrick, J.~E., Hook, I.~M., Swann, E., {et~al.} 2021, Monthly Notices of the
  Royal Astronomical Society, \dodoi{10.1093/mnras/stab2343}

\bibitem[{Caswell {et~al.}(2020)Caswell, Droettboom, Lee, Hunter, Andrade,
  Firing, Hoffmann, Klymak, Stansby, Varoquaux, Nielsen, Root, May, Elson,
  Seppänen, Dale, {Jae-Joon Lee}, McDougall, Straw, Hobson, Gohlke, Yu, Ma,
  Vincent, Silvester, Moad, Kniazev, {, Hannah}, Ernest, \&
  Ivanov}]{matplotlib2020}
Caswell, T.~A., Droettboom, M., Lee, A., {et~al.} 2020, matplotlib/matplotlib:
  REL: v3.3.2,  Zenodo, \dodoi{10.5281/ZENODO.4030140}

\bibitem[{Charnock \& Moss(2017)}]{Charnock2017}
Charnock, T., \& Moss, A. 2017, The Astrophysical Journal, 837, L28,
  \dodoi{10.3847/2041-8213/aa603d}

\bibitem[{Chen {et~al.}(2013)Chen, Qian, \& Meng}]{Chen2013}
Chen, N., Qian, Z., \& Meng, X. 2013, Mathematical Problems in Engineering,
  2013, 1, \dodoi{10.1155/2013/461983}

\bibitem[{Friedman(2001)}]{Friedman2001}
Friedman, J.~H. 2001, The Annals of Statistics, 29, 1189,
  \dodoi{10.1214/aos/1013203451}

\bibitem[{Gonzalez {et~al.}(2018)Gonzalez, Clarkson, Debattista, Johnson, Rich,
  Bono, Dall'Ora, Gizis, Kallivayalil, Kawata, Lucas, Minniti, Schiavon,
  Strader, Street, Valenti, \& Zoccali}]{Gonzalez2018}
Gonzalez, O.~A., Clarkson, W., Debattista, V.~P., {et~al.} 2018, arXiv preprint
  arXiv:1812.08670

\bibitem[{Guillochon {et~al.}(2017)Guillochon, Parrent, Kelley, \&
  Margutti}]{Guillochon2017}
Guillochon, J., Parrent, J., Kelley, L.~Z., \& Margutti, R. 2017, The
  Astrophysical Journal, 835, 64, \dodoi{10.3847/1538-4357/835/1/64}

\bibitem[{Harris {et~al.}(2020)Harris, Millman, van~der Walt, Gommers,
  Virtanen, Cournapeau, Wieser, Taylor, Berg, Smith, Kern, Picus, Hoyer, van
  Kerkwijk, Brett, Haldane, del R{'{\i}}o, Wiebe, Peterson,
  G{'{e}}rard-Marchant, Sheppard, Reddy, Weckesser, Abbasi, Gohlke, \&
  Oliphant}]{harris2020array}
Harris, C.~R., Millman, K.~J., van~der Walt, S.~J., {et~al.} 2020, Nature, 585,
  357, \dodoi{10.1038/s41586-020-2649-2}

\bibitem[{Hlo{\v{z}}ek {et~al.}(2020)Hlo{\v{z}}ek, Ponder, Malz, Dai, Narayan,
  Ishida, Jr, Bahmanyar, Biswas, Galbany, Jha, Jones, Kessler, Lochner,
  Mahabal, Mandel, Martínez-Galarza, McEwen, Muthukrishna, Peiris, Peters, \&
  Setzer}]{ResultsPlasticc2020}
Hlo{\v{z}}ek, R., Ponder, K.~A., Malz, A.~I., {et~al.} 2020, arXiv preprint
  arXiv:2012.12392

\bibitem[{Hotelling(1933)}]{Hotelling1933}
Hotelling, H. 1933, Journal of Educational Psychology, 24, 417,
  \dodoi{10.1037/h0071325}

\bibitem[{Hunter(2007)}]{matplotlib2007}
Hunter, J.~D. 2007, Computing in Science \& Engineering, 9, 90,
  \dodoi{10.1109/MCSE.2007.55}

\bibitem[{Istas(1992)}]{Istas1992}
Istas, J. 1992, Annales de l'I.H.P. Probabilités et statistiques, 28, 537.
\newblock \url{http://www.numdam.org/item/AIHPB_1992__28_4_537_0/}

\bibitem[{Ivezi{\'c} {et~al.}(2018)Ivezi{\'c}, Jones, \&
  Ribeiro}]{ivezic2018call}
Ivezi{\'c}, {\v{Z}}., Jones, L., \& Ribeiro, T. 2018, {Call for White Papers on
  LSST Cadence Optimization}

\bibitem[{Ivezi{\'{c}} {et~al.}(2019)Ivezi{\'{c}}, Kahn, Tyson, Abel, Acosta,
  Allsman, Alonso, AlSayyad, Anderson, Andrew, Angel, Angeli, Ansari,
  Antilogus, Araujo, Armstrong, Arndt, Astier, Aubourg, Auza, Axelrod, Bard,
  Barr, Barrau, Bartlett, Bauer, Bauman, Baumont, Bechtol, Bechtol, Becker,
  Becla, Beldica, Bellavia, Bianco, Biswas, Blanc, Blazek, Blandford, Bloom,
  Bogart, Bond, Booth, Borgland, Borne, Bosch, Boutigny, Brackett, Bradshaw,
  Brandt, Brown, Bullock, Burchat, Burke, Cagnoli, Calabrese, Callahan, Callen,
  Carlin, Carlson, Chandrasekharan, Charles-Emerson, Chesley, Cheu, Chiang,
  Chiang, Chirino, Chow, Ciardi, Claver, Cohen-Tanugi, Cockrum, Coles,
  Connolly, Cook, Cooray, Covey, Cribbs, Cui, Cutri, Daly, Daniel, Daruich,
  Daubard, Daues, Dawson, Delgado, Dellapenna, de~Peyster, de~Val-Borro, Digel,
  Doherty, Dubois, Dubois-Felsmann, Durech, Economou, Eifler, Eracleous,
  Emmons, Neto, Ferguson, Figueroa, Fisher-Levine, Focke, Foss, Frank, Freemon,
  Gangler, Gawiser, Geary, Gee, Geha, Gessner, Gibson, Gilmore, Glanzman,
  Glick, Goldina, Goldstein, Goodenow, Graham, Gressler, Gris, Guy, Guyonnet,
  Haller, Harris, Hascall, Haupt, Hernandez, Herrmann, Hileman, Hoblitt,
  Hodgson, Hogan, Howard, Huang, Huffer, Ingraham, Innes, Jacoby, Jain, Jammes,
  Jee, Jenness, Jernigan, Jevremovi{\'{c}}, Johns, Johnson, Johnson, Jones,
  Juramy-Gilles, Juri{\'{c}}, Kalirai, Kallivayalil, Kalmbach, Kantor, Karst,
  Kasliwal, Kelly, Kessler, Kinnison, Kirkby, Knox, Kotov, Krabbendam,
  Krughoff, Kub{\'{a}}nek, Kuczewski, Kulkarni, Ku, Kurita, Lage, Lambert,
  Lange, Langton, Guillou, Levine, Liang, Lim, Lintott, Long, Lopez, Lotz,
  Lupton, Lust, MacArthur, Mahabal, Mandelbaum, Markiewicz, Marsh, Marshall,
  Marshall, May, McKercher, McQueen, Meyers, Migliore, Miller, Mills, Miraval,
  Moeyens, Moolekamp, Monet, Moniez, Monkewitz, Montgomery, Morrison, Mueller,
  Muller, Arancibia, Neill, Newbry, Nief, Nomerotski, Nordby, O'Connor, Oliver,
  Olivier, Olsen, O'Mullane, Ortiz, Osier, Owen, Pain, Palecek, Parejko,
  Parsons, Pease, Peterson, Peterson, Petravick, Petrick, Petry, Pierfederici,
  Pietrowicz, Pike, Pinto, Plante, Plate, Plutchak, Price, Prouza, Radeka,
  Rajagopal, Rasmussen, Regnault, Reil, Reiss, Reuter, Ridgway, Riot, Ritz,
  Robinson, Roby, Roodman, Rosing, Roucelle, Rumore, Russo, Saha, Sassolas,
  Schalk, Schellart, Schindler, Schmidt, Schneider, Schneider, Schoening,
  Schumacher, Schwamb, Sebag, Selvy, Sembroski, Seppala, Serio, Serrano, Shaw,
  Shipsey, Sick, Silvestri, Slater, Smith, Smith, Sobhani, Soldahl,
  Storrie-Lombardi, Stover, Strauss, Street, Stubbs, Sullivan, Sweeney,
  Swinbank, Szalay, Takacs, Tether, Thaler, Thayer, Thomas, Thornton, Thukral,
  Tice, Trilling, Turri, Berg, Berk, Vetter, Virieux, Vucina, Wahl, Walkowicz,
  Walsh, Walter, Wang, Wang, Warner, Wiecha, Willman, Winters, Wittman, Wolff,
  Wood-Vasey, Wu, Xin, Yoachim, \& Zhan}]{Ivezic2019}
Ivezi{\'{c}}, {\v{Z}}., Kahn, S.~M., Tyson, J.~A., {et~al.} 2019, The
  Astrophysical Journal, 873, 111, \dodoi{10.3847/1538-4357/ab042c}

\bibitem[{Jones {et~al.}(2017)Jones, Scolnic, Riess, Kessler, Rest, Kirshner,
  Berger, Ortega, Foley, Chornock, Challis, Burgett, Chambers, Draper,
  Flewelling, Huber, Kaiser, Kudritzki, Metcalfe, Wainscoat, \&
  Waters}]{Jones2017}
Jones, D.~O., Scolnic, D.~M., Riess, A.~G., {et~al.} 2017, The Astrophysical
  Journal, 843, 6, \dodoi{10.3847/1538-4357/aa767b}

\bibitem[{Jones {et~al.}(2018)Jones, Scolnic, Riess, Rest, Kirshner, Berger,
  Kessler, Pan, Foley, Chornock, Ortega, Challis, Burgett, Chambers, Draper,
  Flewelling, Huber, Kaiser, Kudritzki, Metcalfe, Tonry, Wainscoat, Waters,
  Gall, Kotak, McCrum, Smartt, \& Smith}]{Jones2018}
---. 2018, The Astrophysical Journal, 857, 51, \dodoi{10.3847/1538-4357/aab6b1}

\bibitem[{Jones {et~al.}(2020)Jones, Yoachim, Ivezic, Neilsen, \&
  Ribeiro}]{Jones2020}
Jones, R.~L., Yoachim, P., Ivezic, Z., Neilsen, E.~H., \& Ribeiro, T. 2020,
  {Survey Strategy and Cadence Choices for the Vera C. Rubin Observatory Legacy
  Survey of Space and Time (LSST)}, Tech. rep., \dodoi{10.5281/ZENODO.4048838}

\bibitem[{Ke {et~al.}(2017)Ke, Meng, Finley, Wang, Chen, Ma, Ye, \&
  Liu}]{ke2017lightgbm}
Ke, G., Meng, Q., Finley, T., {et~al.} 2017, in Advances in Neural Information
  Processing Systems 30, ed. I.~Guyon, U.~V. Luxburg, S.~Bengio, H.~Wallach,
  R.~Fergus, S.~Vishwanathan, \& R.~Garnett, Vol.~30 (Curran Associates, Inc.),
  3146--3154

\bibitem[{Kessler {et~al.}(2010{\natexlab{a}})Kessler, Conley, Jha, \&
  Kuhlmann}]{Kessler2010spcc}
Kessler, R., Conley, A., Jha, S., \& Kuhlmann, S. 2010{\natexlab{a}}, arXiv
  preprint arXiv:1001.5210

\bibitem[{Kessler \& Scolnic(2017)}]{Kessler2017}
Kessler, R., \& Scolnic, D. 2017, The Astrophysical Journal, 836, 56,
  \dodoi{10.3847/1538-4357/836/1/56}

\bibitem[{Kessler {et~al.}(2009)Kessler, Becker, Cinabro, Vanderplas, Frieman,
  Marriner, Davis, Dilday, Holtzman, Jha, Lampeitl, Sako, Smith, Zheng, Nichol,
  Bassett, Bender, Depoy, Doi, Elson, Filippenko, Foley, Garnavich, Hopp,
  Ihara, Ketzeback, Kollatschny, Konishi, Marshall, McMillan, Miknaitis,
  Morokuma, Mörtsell, Pan, Prieto, Richmond, Riess, Romani, Schneider,
  Sollerman, Takanashi, Tokita, van~der Heyden, Wheeler, Yasuda, \&
  York}]{Kessler2009}
Kessler, R., Becker, A.~C., Cinabro, D., {et~al.} 2009, The Astrophysical
  Journal Supplement Series, 185, 32, \dodoi{10.1088/0067-0049/185/1/32}

\bibitem[{Kessler {et~al.}(2010{\natexlab{b}})Kessler, Bassett, Belov,
  Bhatnagar, Campbell, Conley, Frieman, Glazov, Gonz{\'{a}}lez-Gait{\'{a}}n,
  Hlozek, Jha, Kuhlmann, Kunz, Lampeitl, Mahabal, Newling, Nichol, Parkinson,
  Philip, Poznanski, Richards, Rodney, Sako, Schneider, Smith, Stritzinger, \&
  Varughese}]{Kessler2010results}
Kessler, R., Bassett, B., Belov, P., {et~al.} 2010{\natexlab{b}}, Publications
  of the Astronomical Society of the Pacific, 122, 1415, \dodoi{10.1086/657607}

\bibitem[{Kessler {et~al.}(2019)Kessler, Narayan, Avelino, Bachelet, Biswas,
  Brown, Chernoff, Connolly, Dai, Daniel, Stefano, Drout, Galbany,
  Gonz{\'{a}}lez-Gait{\'{a}}n, Graham, Hlo{\v{z}}ek, Ishida, Guillochon, Jha,
  Jones, Mandel, Muthukrishna, O'Grady, Peters, Pierel, Ponder, Pr{\v{s}}a,
  Rodney, \& and}]{Kessler2019}
Kessler, R., Narayan, G., Avelino, A., {et~al.} 2019, Publications of the
  Astronomical Society of the Pacific, 131, 094501,
  \dodoi{10.1088/1538-3873/ab26f1}

\bibitem[{Kluyver {et~al.}(2016)Kluyver, Ragan-Kelley, P{\'e}rez, Granger,
  Bussonnier, Frederic, Kelley, Hamrick, Grout, Corlay, Ivanov, Avila, Abdalla,
  Willing, \& development team}]{jupyter2016}
Kluyver, T., Ragan-Kelley, B., P{\'e}rez, F., {et~al.} 2016, in Positioning and
  Power in Academic Publishing: Players, Agents and Agendas, ed. F.~Loizides \&
  B.~Scmidt (Netherlands: IOS Press), 87--90.
\newblock \url{https://eprints.soton.ac.uk/403913/}

\bibitem[{Krekel {et~al.}(2004)Krekel, Oliveira, Pfannschmidt, Bruynooghe,
  Laugher, \& Bruhin}]{pytest6.2.2}
Krekel, H., Oliveira, B., Pfannschmidt, R., {et~al.} 2004, pytest 6.2.2.
\newblock \url{https://github.com/pytest-dev/pytest}

\bibitem[{Kunz {et~al.}(2007)Kunz, Bassett, \& Hlozek}]{Kunz2007}
Kunz, M., Bassett, B.~A., \& Hlozek, R. 2007, Physical Review D, 75, 103508,
  \dodoi{10.1103/PhysRevD.75.103508}

\bibitem[{Laine {et~al.}(2018)Laine, Martinez-Delgado, Trujillo, Duc,
  Grillmair, Frenk, Hendel, Johnston, Mihos, Moustakas, Beaton, Romanowsky,
  Greco, \& Erkal}]{Laine2018}
Laine, S., Martinez-Delgado, D., Trujillo, I., {et~al.} 2018, arXiv preprint
  arXiv:1812.04897

\bibitem[{Lee {et~al.}(2019{\natexlab{a}})Lee, Gommers, Waselewski, Wohlfahrt,
  \& O'Leary}]{Lee2019}
Lee, G., Gommers, R., Waselewski, F., Wohlfahrt, K., \& O'Leary, A.
  2019{\natexlab{a}}, Journal of Open Source Software, 4, 1237,
  \dodoi{10.21105/joss.01237}

\bibitem[{Lee {et~al.}(2019{\natexlab{b}})Lee, Gommers, Wohlfahrt, Wasilewski,
  O'Leary, Nahrstaedt, Hurtado, Sauvé, Arildsen, Oliveira, Pelt, Agrawal,
  {SylvainLan}, Pelletier, Brett, Yu, {Saket Choudhary}, Tricoli, Craig,
  {Lokesh Ravindranathan}, Dan, {Jakirkham}, Antonello, Laszuk, Goertzen,
  Goldberg, {Balint Reczey}, {0-Tree}, {Arfon Smith}, \& {Asnt}}]{Lee2019a}
Lee, G.~R., Gommers, R., Wohlfahrt, K., {et~al.} 2019{\natexlab{b}},
  PyWavelets/pywt: PyWavelets 1.1.1,  Zenodo, \dodoi{10.5281/ZENODO.3510098}

\bibitem[{Lochner {et~al.}(2013)Lochner, Bassett, Varughese, Hlozek, Kunz,
  Smith, \& Newling}]{Lochner2013}
Lochner, M., Bassett, B.~A., Varughese, M., {et~al.} 2013, Journal of Cosmology
  and Astroparticle Physics, 2013, 039, \dodoi{10.1088/1475-7516/2013/01/039}

\bibitem[{Lochner {et~al.}(2016)Lochner, McEwen, Peiris, Lahav, \&
  Winter}]{Lochner2016}
Lochner, M., McEwen, J.~D., Peiris, H.~V., Lahav, O., \& Winter, M.~K. 2016,
  The Astrophysical Journal Supplement Series, 225, 31

\bibitem[{Lochner {et~al.}(2018)Lochner, Scolnic, Awan, Regnault, Gris,
  Mandelbaum, Gawiser, Almoubayyed, Setzer, Huber, Graham, Hložek, Biswas,
  Eifler, Rothchild, Jr, Blazek, Chang, Collett, Goobar, Hook, Jarvis, Jha,
  Kim, Marshall, McEwen, Moniez, Newman, Peiris, Petrushevska, Rhodes,
  Sevilla-Noarbe, Slosar, Suyu, Tyson, \& Yoachim}]{DESCObservingStrategy2018}
Lochner, M., Scolnic, D.~M., Awan, H., {et~al.} 2018, arXiv preprint
  arXiv:1812.00515

\bibitem[{Lochner {et~al.}(2021)Lochner, Scolnic, Almoubayyed, Anguita, Awan,
  Gawiser, Gontcho, Gris, Huber, Jha, Jones, Kim, Mandelbaum, Marshall,
  Petrushevska, Regnault, Setzer, Suyu, Yoachim, Biswas, Blaineau, Hook,
  Moniez, Neilsen, Peiris, Rothchild, Stubbs, \& {the LSST Dark Energy Science
  Collaboration}}]{osmetrics}
Lochner, M., Scolnic, D., Almoubayyed, H., {et~al.} 2021, arXiv preprint
  arXiv:2104.05676.
\newblock \doarXiv{2104.05676}

\bibitem[{{LSST Science Collaboration} {et~al.}(2009){LSST Science
  Collaboration}, Abell, Allison, Anderson, Andrew, Angel, Armus, Arnett,
  Asztalos, Axelrod, Bailey, Ballantyne, Bankert, Barkhouse, Barr, Barrientos,
  Barth, Bartlett, Becker, Becla, Beers, Bernstein, Biswas, Blanton, Bloom,
  Bochanski, Boeshaar, Borne, Bradac, Brandt, Bridge, Brown, Brunner, Bullock,
  Burgasser, Burge, Burke, Cargile, Chandrasekharan, Chartas, Chesley, Chu,
  Cinabro, Claire, Claver, Clowe, Connolly, Cook, Cooke, Cooray, Covey,
  Culliton, de~Jong, de~Vries, Debattista, Delgado, Dell'Antonio, Dhital,
  Stefano, Dickinson, Dilday, Djorgovski, Dobler, Donalek, Dubois-Felsmann,
  Durech, Eliasdottir, Eracleous, Eyer, Falco, Fan, Fassnacht, Ferguson,
  Fernandez, Fields, Finkbeiner, Figueroa, Fox, Francke, Frank, Frieman,
  Fromenteau, Furqan, Galaz, Gal-Yam, Garnavich, Gawiser, Geary, Gee, Gibson,
  Gilmore, Grace, Green, Gressler, Grillmair, Habib, Haggerty, Hamuy, Harris,
  Hawley, Heavens, Hebb, Henry, Hileman, Hilton, Hoadley, Holberg, Holman,
  Howell, Infante, Ivezic, Jacoby, Jain, R, Jedicke, Jee, Jernigan, Jha,
  Johnston, Jones, Juric, Kaasalainen, Styliani, Kafka, Kahn, Kaib, Kalirai,
  Kantor, Kasliwal, Keeton, Kessler, Knezevic, Kowalski, Krabbendam, Krughoff,
  Kulkarni, Kuhlman, Lacy, Lepine, Liang, Lien, Lira, Long, Lorenz, Lotz,
  Lupton, Lutz, Macri, Mahabal, Mandelbaum, Marshall, May, McGehee, Meadows,
  Meert, Milani, Miller, Miller, Mills, Minniti, Monet, Mukadam, Nakar, Neill,
  Newman, Nikolaev, Nordby, O'Connor, Oguri, Oliver, Olivier, Olsen, Olsen,
  Olszewski, Oluseyi, Padilla, Parker, Pepper, Peterson, Petry, Pinto, Pizagno,
  Popescu, Prsa, Radcka, Raddick, Rasmussen, Rau, Rho, Rhoads, Richards,
  Ridgway, Robertson, Roskar, Saha, Sarajedini, Scannapieco, Schalk, Schindler,
  Schmidt, Schmidt, Schneider, Schumacher, Scranton, Sebag, Seppala, Shemmer,
  Simon, Sivertz, Smith, Smith, Smith, Spitz, Stanford, Stassun, Strader,
  Strauss, Stubbs, Sweeney, Szalay, Szkody, Takada, Thorman, Trilling, Trimble,
  Tyson, Berg, Berk, VanderPlas, Verde, Vrsnak, Walkowicz, Wandelt, Wang, Wang,
  Warner, Wechsler, West, Wiecha, Williams, Willman, Wittman, Wolff,
  Wood-Vasey, Wozniak, Young, Zentner, \& Zhan}]{LSSTScienceBook2009}
{LSST Science Collaboration}, Abell, P.~A., Allison, J., {et~al.} 2009, arXiv
  preprint arXiv:0912.0201

\bibitem[{{LSST Science Collaboration} {et~al.}(2017){LSST Science
  Collaboration}, Marshall, Anguita, Bianco, Bellm, Brandt, Clarkson, Connolly,
  Gawiser, Ivezic, Jones, Lochner, Lund, Mahabal, Nidever, Olsen, Ridgway,
  Rhodes, Shemmer, Trilling, Vivas, Walkowicz, Willman, Yoachim, Anderson,
  Antilogus, Angus, Arcavi, Awan, Biswas, Bell, Bennett, Britt, Buzasi,
  Casetti-Dinescu, Chomiuk, Claver, Cook, Davenport, Debattista, Digel, Doctor,
  Firth, Foley, fai Fong, Galbany, Giampapa, Gizis, Graham, Grillmair, Gris,
  Haiman, Hartigan, Hawley, Hlozek, Jha, Johns-Krull, Kanbur, Kalogera,
  Kashyap, Kasliwal, Kessler, Kim, Kurczynski, Lahav, Liu, Malz, Margutti,
  Matheson, McEwen, McGehee, Meibom, Meyers, Monet, Neilsen, Newman, O'Dowd,
  Peiris, Penny, Peters, Poleski, Ponder, Richards, Rho, Rubin, Schmidt,
  Schuhmann, Shporer, Slater, Smith, Soares-Santos, Stassun, Strader, Strauss,
  Street, Stubbs, Sullivan, Szkody, Trimble, Tyson, de~Val-Borro, Valenti,
  Wagoner, Wood-Vasey, \& Zauderer}]{LSSTObservingStartegy2017}
{LSST Science Collaboration}, Marshall, P., Anguita, T., {et~al.} 2017, arXiv
  preprint arXiv:1708.04058, \dodoi{10.5281/zenodo.842713}

\bibitem[{MacKay(2003)}]{MacKay2003}
MacKay, D. J.~C. 2003, {I}nformation {T}heory, {I}nference and {L}earning
  {A}lgorithms (Cambridge University Pr.)

\bibitem[{Malz {et~al.}(2019)Malz, Hlo{\v{z}}ek, Allam, Bahmanyar, Biswas, Dai,
  Galbany, Ishida, Jha, Jones, Kessler, Lochner, Mahabal, Mandel,
  Mart{\'{\i}}nez-Galarza, McEwen, Muthukrishna, Narayan, Peiris, Peters,
  Ponder, \& and}]{malz2018metric}
Malz, A.~I., Hlo{\v{z}}ek, R., Allam, T., {et~al.} 2019, The Astronomical
  Journal, 158, 171, \dodoi{10.3847/1538-3881/ab3a2f}

\bibitem[{Mockus {et~al.}(1978)Mockus, Tiesis, \& Zilinskas}]{Mockus1978}
Mockus, J., Tiesis, V., \& Zilinskas, A. 1978, Towards Global Optimization, 2,
  2

\bibitem[{Muthukrishna {et~al.}(2019)Muthukrishna, Narayan, Mandel, Biswas, \&
  Hlo{\v{z}}ek}]{Muthukrishna2019}
Muthukrishna, D., Narayan, G., Mandel, K.~S., Biswas, R., \& Hlo{\v{z}}ek, R.
  2019, Publications of the Astronomical Society of the Pacific, 131, 118002,
  \dodoi{10.1088/1538-3873/ab1609}

\bibitem[{Narayan {et~al.}(2018)Narayan, Zaidi, Soraisam, Wang, Lochner,
  Matheson, Saha, Yang, Zhao, Kececioglu, Scheidegger, Snodgrass, Axelrod,
  Jenness, Maier, Ridgway, Seaman, Evans, Singh, Taylor, Toeniskoetter, Welch,
  \& and}]{Narayan2018}
Narayan, G., Zaidi, T., Soraisam, M.~D., {et~al.} 2018, The Astrophysical
  Journal Supplement Series, 236, 9, \dodoi{10.3847/1538-4365/aab781}

\bibitem[{pandas~development team(2020)}]{reback2020pandas}
pandas~development team, T. 2020, pandas-dev/pandas: Pandas, latest,  Zenodo,
  \dodoi{10.5281/zenodo.3509134}

\bibitem[{Pasquet {et~al.}(2019)Pasquet, Pasquet, Chaumont, \&
  Fouchez}]{Pasquet2019}
Pasquet, J., Pasquet, J., Chaumont, M., \& Fouchez, D. 2019, Astronomy {\&}
  Astrophysics, 627, A21, \dodoi{10.1051/0004-6361/201834473}

\bibitem[{Pearson(1901)}]{Pearson1901}
Pearson, K. 1901, The London, Edinburgh, and Dublin Philosophical Magazine and
  Journal of Science, 2, 559, \dodoi{10.1080/14786440109462720}

\bibitem[{Pedregosa {et~al.}(2011)Pedregosa, Varoquaux, Gramfort, Michel,
  Thirion, Grisel, Blondel, Prettenhofer, Weiss, Dubourg, Vanderplas, Passos,
  Cournapeau, Brucher, Perrot, \& Duchesnay}]{scikit-learn2011}
Pedregosa, F., Varoquaux, G., Gramfort, A., {et~al.} 2011, Journal of Machine
  Learning Research, 12, 2825

\bibitem[{{PLAsTiCC Modelers}(2019)}]{PLAsTiCCModelers2019}
{PLAsTiCC Modelers}. 2019, Libraries \& Recommended Citations for using
  PLAsTiCC Models,  Zenodo, \dodoi{10.5281/ZENODO.2612896}

\bibitem[{{PLAsTiCC Team} \& {PLAsTiCC Modelers}(2019)}]{PLASTICCTeam2019}
{PLAsTiCC Team}, \& {PLAsTiCC Modelers}. 2019, Unblinded Data for {PLAsTiCC}
  Classification Challenge,  Zenodo, \dodoi{10.5281/ZENODO.2535746}

\bibitem[{Pope(2019)}]{Pope2019}
Pope, C.~A. 2019, PhD thesis, University of Leeds.
\newblock \url{https://etheses.whiterose.ac.uk/25066/}

\bibitem[{Rasmussen \& {W}illiams(2005)}]{Rasmussen2005}
Rasmussen, C.~E., \& {W}illiams, C. K.~I. 2005, {G}aussian {P}rocesses for
  {M}achine {L}earning (MIT Press Ltd)

\bibitem[{Revsbech {et~al.}(2017)Revsbech, Trotta, \& van Dyk}]{Revsbech2017}
Revsbech, E.~A., Trotta, R., \& van Dyk, D.~A. 2017, Monthly Notices of the
  Royal Astronomical Society, 473, 3969, \dodoi{10.1093/mnras/stx2570}

\bibitem[{Riess {et~al.}(1998)Riess, Filippenko, Challis, Clocchiatti, Diercks,
  Garnavich, Gilliland, Hogan, Jha, Kirshner, Leibundgut, Phillips, Reiss,
  Schmidt, Schommer, Smith, Spyromilio, Stubbs, Suntzeff, \& Tonry}]{Riess1998}
Riess, A.~G., Filippenko, A.~V., Challis, P., {et~al.} 1998, The Astronomical
  Journal, 116, 1009, \dodoi{10.1086/300499}

\bibitem[{Roberts {et~al.}(2017)Roberts, Lochner, Fonseca, Bassett, Lablanche,
  \& Agarwal}]{Roberts2017}
Roberts, E., Lochner, M., Fonseca, J., {et~al.} 2017, Journal of Cosmology and
  Astroparticle Physics, 2017, 036, \dodoi{10.1088/1475-7516/2017/10/036}

\bibitem[{Scolnic {et~al.}(2018)Scolnic, Lochner, Gris, Regnault, Hložek,
  Aldering, Jr, Awan, Biswas, Blazek, Chang, Gawiser, Goobar, Hook, Jha,
  McEwen, Mandelbaum, Marshall, Neilsen, Rhodes, Rothchild, Noarbe, Slosar, \&
  Yoachim}]{DESCObservingStrategy2018ddf}
Scolnic, D.~M., Lochner, M., Gris, P., {et~al.} 2018, arXiv preprint
  arXiv:1812.00516

\bibitem[{Snoek {et~al.}(2012)Snoek, Larochelle, \& Adams}]{snoek2012}
Snoek, J., Larochelle, H., \& Adams, R.~P. 2012, arXiv preprint arXiv:1206.2944

\bibitem[{Sooknunan {et~al.}(2021)Sooknunan, Lochner, Bassett, Peiris, Fender,
  Stewart, Pietka, Woudt, McEwen, \& Lahav}]{Sooknunan2021}
Sooknunan, K., Lochner, M., Bassett, B.~A., {et~al.} 2021, Monthly Notices of
  the Royal Astronomical Society, 502, 206, \dodoi{10.1093/mnras/staa3873}

\bibitem[{Swann {et~al.}(2019)Swann, Sullivan, Carrick, Hoenig, Hook, Kotak,
  Maguire, Nichol, \& Smartt}]{Swann2019}
Swann, E., Sullivan, M., Carrick, J., {et~al.} 2019, The Messenger vol. 175,
  pp. 58-61, March 2019., \dodoi{10.18727/0722-6691/5129}

\bibitem[{Takahashi {et~al.}(2020)Takahashi, Suzuki, Yasuda, Kimura, Ueda,
  Tanaka, Tominaga, \& Yoshida}]{Takahashi2020}
Takahashi, I., Suzuki, N., Yasuda, N., {et~al.} 2020, Publications of the
  Astronomical Society of Japan, 72, \dodoi{10.1093/pasj/psaa082}

\bibitem[{{The Dark Energy Survey Collaboration} \& Flaugher(2005)}]{DES2005}
{The Dark Energy Survey Collaboration}, \& Flaugher, B. 2005, International
  Journal of Modern Physics A, 20, 3121, \dodoi{10.1142/s0217751x05025917}

\bibitem[{{The PLAsTiCC team} {et~al.}(2018){The PLAsTiCC team}, Allam~Jr,
  Bahmanyar, Biswas, Dai, Galbany, Hlo{\v{z}}ek, Ishida, Jha, Jones, Kessler,
  {et~al.}}]{allam2018photometric}
{The PLAsTiCC team}, Allam~Jr, T., Bahmanyar, A., {et~al.} 2018, arXiv preprint
  arXiv:1810.00001

\bibitem[{Van~Rossum(2020)}]{van1995python}
Van~Rossum, G. 2020, The Python Library Reference, release 3.8.2 (Python
  Software Foundation)

\bibitem[{Varughese {et~al.}(2015)Varughese, von Sachs, Stephanou, \&
  Bassett}]{Varughese2015}
Varughese, M.~M., von Sachs, R., Stephanou, M., \& Bassett, B.~A. 2015, Monthly
  Notices of the Royal Astronomical Society, 453, 2849,
  \dodoi{10.1093/mnras/stv1816}

\bibitem[{Villar {et~al.}(2020)Villar, Hosseinzadeh, Berger, Ntampaka, Jones,
  Challis, Chornock, Drout, Foley, Kirshner, Lunnan, Margutti, Milisavljevic,
  Sanders, Pan, Rest, Scolnic, Magnier, Metcalfe, Wainscoat, \&
  Waters}]{Villar2020}
Villar, V.~A., Hosseinzadeh, G., Berger, E., {et~al.} 2020, The Astrophysical
  Journal, 905, 94, \dodoi{10.3847/1538-4357/abc6fd}

\bibitem[{Virtanen {et~al.}(2020)Virtanen, Gommers, Oliphant, Haberland, Reddy,
  Cournapeau, Burovski, Peterson, Weckesser, Bright, {van der Walt}, Brett,
  Wilson, Millman, Mayorov, Nelson, Jones, Kern, Larson, Carey, Polat, Feng,
  Moore, {VanderPlas}, Laxalde, Perktold, Cimrman, Henriksen, Quintero, Harris,
  Archibald, Ribeiro, Pedregosa, {van Mulbregt}, \& {SciPy 1.0
  Contributors}}]{2020SciPy}
Virtanen, P., Gommers, R., Oliphant, T.~E., {et~al.} 2020, Nature Methods, 17,
  261, \dodoi{10.1038/s41592-019-0686-2}

\bibitem[{Waskom {et~al.}(2020)Waskom, Botvinnik, Gelbart, Ostblom, Hobson,
  Lukauskas, Gemperline, Augspurger, Halchenko, Warmenhoven, Cole, Ruiter,
  Vanderplas, Hoyer, Pye, Miles, {Corban Swain}, Meyer, Martin, Bachant,
  Quintero, Kunter, Villalba, {, Brian}, Fitzgerald, {C.G. Evans}, Williams,
  O'Kane, Yarkoni, \& Brunner}]{Waskom2020seabor}
Waskom, M., Botvinnik, O., Gelbart, M., {et~al.} 2020, mwaskom/seaborn: v0.11.0
  (Sepetmber 2020),  Zenodo, \dodoi{10.5281/ZENODO.4019146}

\bibitem[{{W}es {M}c{K}inney(2010)}]{mckinney2010pandas}
{W}es {M}c{K}inney. 2010, in {P}roceedings of the 9th {P}ython in {S}cience
  {C}onference, ed. {S}t\'efan van~der {W}alt \& {J}arrod {M}illman, 56 -- 61,
  \dodoi{10.25080/Majora-92bf1922-00a}

\bibitem[{Zhang {et~al.}(2017)Zhang, Si, \& Hsieh}]{lgbm2017}
Zhang, H., Si, S., \& Hsieh, C.-J. 2017, arXiv preprint arXiv:1706.08359

\end{thebibliography}

\end{document}